\documentclass[fullpage, 10pts]{article}
\usepackage{fullpage}
\usepackage{authblk}
\usepackage{graphicx}
\usepackage{epstopdf, epsfig}
\usepackage{tikz}
\usepackage{hyperref}
\usepackage[sort&compress,square,comma,authoryear]{natbib}
\usepackage{amsmath}
\usepackage{amssymb}

\title{A note on Stokes' problem in dense granular media using the $\mu(I)$--rheology}

\author[1]{J. John Soundar Jerome\footnote{\href{john-soundar@univ-lyon1.fr}{john-soundar@univ-lyon1.fr}}}
\author[1]{B. Di Pierro\footnote{\href{bastien.di-pierro@univ-lyon1.fr}{bastien.di-pierro@univ-lyon1.fr}}}

\affil[1]{Universit\'{e} de Lyon, Universit\'{e} Claude Bernard Lyon $1$, Laboratoire de M\'{e}canique des Fluides et d'Acoustique, CNRS UMR--$5509$, Boulevard $11$ novembre, F--$69622$ Villeurbanne cedex, LYON, France}

\begin{document}

\maketitle

\begin{abstract}
The classical Stokes' problem describing the fluid motion due to a steadily moving infinite wall is revisited in the context of dense granular flows of mono-dispersed beads using the recently proposed $\mu(I)$--rheology. In Newtonian fluids, molecular diffusion brings about a self-similar velocity profile and the boundary layer in which the fluid motion takes place increases indefinitely with time $t$ as $\sqrt{\nu t}$, where $\nu$ is the kinematic viscosity. For a dense granular visco-plastic liquid, it is shown that the local shear stress, when properly rescaled, exhibits self-similar behaviour at short-time scales and it then rapidly evolves towards a steady-state solution. The resulting shear layer increases in thickness as $\sqrt{\nu_g t}$ analogous to a Newtonian fluid where $\nu_g$ is an equivalent granular kinematic viscosity depending not only on the intrinsic properties of the granular media such as grain diameter $d$, density $\rho$ and friction coefficients but also on the applied pressure $p_w$ at the moving wall and the solid fraction $\phi$ (constant). In addition, the $\mu(I)$--rheology indicates that this growth continues until reaching the steady-state boundary layer thickness $\delta_s = \beta_w (p_w/\phi \rho g )$, independent of the grain size, at about a finite time proportional to $\beta_w^2 (p_w/\rho g d)^{3/2} \sqrt{d/g}$, where $g$ is the acceleration due to gravity and $\beta_w = (\tau_w - \tau_s)/\tau_s$ is the relative surplus of the steady-state wall shear-stress $\tau_w$ over the critical wall shear stress $\tau_s$ (yield stress) that is needed to bring the granular media into motion. For the case of Stokes' first problem when the wall shear stress $\tau_w$ is imposed externally, the $\mu(I)$--rheology suggests that the wall velocity simply grows as $\sqrt{t}$ before saturating to a constant value whereby the internal resistance of the granular media balances out the applied stresses. In contrast, for the case with an externally imposed wall speed $u_w$, the dense granular media near the wall initially maintains a shear stress very close to $\tau_d$ which is the maximum internal resistance via grain-grain contact friction within the context of the $\mu(I)$--rheology. Then the wall shear stress $\tau_w$ decreases as $1/\sqrt{t}$ until ultimately saturating to a constant value so that it gives precisely the same steady state solution as for the imposed shear stress case. Thereby, the steady-state wall velocity, wall shear stress and the applied wall pressure are related as $u_w \sim  ({g \delta_s^2}/{\nu_g}) f( \beta_w )$ where $f( \beta_w )$ is either $\mathcal{O}(1)$ if $\tau_w \sim \tau_s$ or logarithmically large as $\tau_w$ approaches $\tau_d$. 

\end{abstract}

%\begin{keywords}
%Authors should not enter keywords on the manuscript, as these must be chosen by the author during the online submission process and will then be added during the typesetting process (see http://journals.cambridge.org/data/\linebreak[3]relatedlink/jfm-\linebreak[3]keywords.pdf for the full list)
%\end{keywords}

\section {Introduction}
%%%%%%%%%%%%%%%%%%%%%%%%%%%%%%%%%%%%
\label{sec:Intro}
Flowing matter containing a dense collection of grains like sand, gravel, cereals, sugar, etc. is ubiquitous in nature as well as in many industrial processes. Such granular media exist in various states at any given common flow situation depending strongly on the energy supplied by external deformation and/or shear stresses \citep{jaeger1996granular}. And so, they show a very rich phenomenology \citep{liu1998nonlinear, aranson2006patterns, gray2003shock} : a gaseous regime wherein the flow is very rapid and dilute, and the particles interact by \textit{collision} \citep{jenkins1983theory} and a quasi-static regime in which the material deformation is extremely slow wherein \textit{frictional contacts} between particles dominate the rheology as often in soil mechanics \citep{hutter1994flows}. Indeed, there exists an intermediate regime in the presence of both collisions and friction that result in huge dissipation. Here a dense granular media behaves like a viscoplastic liquid \citep{forterre2008flows, Andreotti_Forterre_Pouliquen_2011}. A decade ago, generalising the scalar rheology of \citet{midi2004dense}, \citet{jop2006constitutive} proposed the so-called $\mu(I)$--rheology to describe such a dense granular liquid state.  It has since been well-exploited often via direct numerical simulations to study and model many a common flow configurations \citep{kamrin2010nonlinear, cawthorn2011thesis, lagree2011granular, staron2012granular, chauchat2014three, gray2014depth, baker2016two}.

However, recent works by \citet{barker2015well, martin2017continuum, goddard2017stability} illustrate that the governing equations of the $\mu(I)$--rheology can exhibit \textit{ill-posed} behaviour in the parameter range corresponding to quasi-gaseous and quasi-static regimes, respectively. \citet{joseph1990short} showed that ill-posed problems suffer from the so-called \textit{Hadamard} instability and so, they characterized ill-posedness through a stability analysis that identifies exponential temporal growth of short-wavelength perturbations. And as a consequence, grid-dependent numerical results may not converge as the spatial refinement is enhanced for these cases \citep[see][p. 224]{joseph1990short}. In particular, \citet{barker2015well} demonstrated both theoretically and numerically the governing equations of the $\mu(I)$--rheology are \textit{Hadamard} unstable even for the simple case of Bagnold flow. More recently, \citet{martin2017continuum} also observed it in their numerical simulations for the case of granular column collapse on inclined channels. Nonetheless novel attempts to regularise the governing equations via a proper functional form of $\mu(I)$, atleast in the quasi-static regime, have already been proposed by \citet{barker2017well, barker2017partial}. They successfully simulated granular roll-waves in two dimensions and it now remains to see if their regularisation is valid also in direct computations of other unsteady granular flows.

\begin{figure}
\begin{center}
\epsfig{file=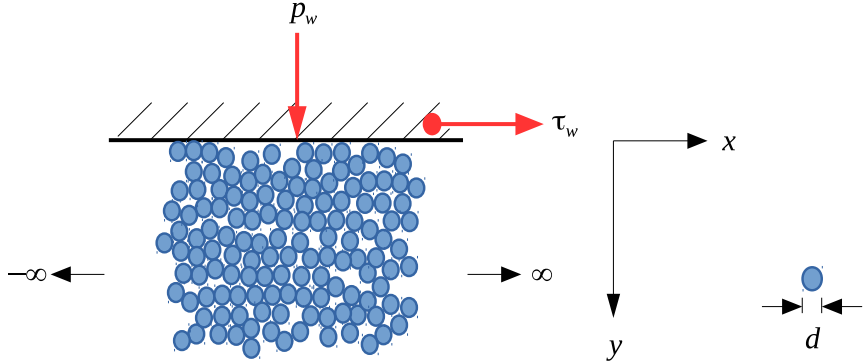,width=1\textwidth,keepaspectratio=true}
\end{center}
\caption{Schematic of the problem : an impulsively started, infinitely long flat plate over a semi-infinite dense granular media consisting of mono-dispersed spherical grains.}
\label{fig:Schematic}
\end{figure}

In this context, this work aims to determine, both numerically and theoretically, the time evolution characteristics of an unsteady, non-uniform velocity and shear stress fields arising in the $\mu(I)$--rheology for a canonical flow situation, namely, the so-called Stokes' first problem \citep{stokes1851effect} of the fluid motion that is brought about by impulsively starting an infinite wall. Unlike the classical case, the granular media is placed underneath the plate (see figure \ref{fig:Schematic}). It is the simplest unsteady parallel flow in which some important features of fluid flows such as transverse momentum transfer and the resulting boundary layer development due to direct balance between local fluid acceleration and the friction forces can be treated. It is also known as the \textit{dragged-plate} problem in \citet{cawthorn2011thesis, Andreotti_Forterre_Pouliquen_2011} where only a steady-state analytical solution can so far been found. The objective of the present work is to treat the transient solution and its characteristics.

Note that there has been a steady interest in Stokes's first problem for non-Newtonian fluids, in particular, visco-elastic fluids \citep[see][]{morrison1956wave, tanner1962note, preziosi1987stokes, devakar2008stokes, devakar2009stokes}. Similarly, Stokes' second problem \citep{schlichting1968boundary, panton1968transient} that considers the time evolution of the velocity field due to an horizontally oscillating infinite flat plate is recently studied for viscoelastic fluids by \citet{devakar2008stokes}. Its applications include high-frequency microfluidics \citep{YakhotColosqui_JFM2007, YakhotOscillatingFlows_PRL2008} for viscoelastic materials and rheometers for viscoplastic \citep{Balforth_JNonNewt2009} and power law fluids \citep{Pritchard_JNonNewt2011}. Recent literature also considers a third type of Stokes' problem wherein a transient velocity field is set-up by suddenly applying a body force to the fluid that is initially at rest. In fact, for a granular media, \citet{Jop2007Stokes3rdProb} used this configuration to numerically validate their proposed $\mu(I)$--rheology. This has later been referred to as Stokes' third problem by \citet{Ancey_JNonNewt2017} for a Herschel--Bulkley material and interestingly, for the numerical resolution, the authors resorted to a Stefan problem, with a moving interface (boundary condition) that separates the sheared and unsheared regions.

The constitutive laws for many non-Newtonian fluids are often non-linear but they can be simplified in the case of Stokes' problems. Whereas the yield stress of the granular material varies in space since the $\mu(I)$--rheology proposes a constitutive law for dense granular flows wherein the media behaves like a visco-plastic liquid with the local viscosity non-linearly related to the local strain--rate as well as the local pressure. Moreover, care should be taken to express a well-posed initial-value problem using the $\mu(I)$--rheology to avoid \textit{Hadamard} instability. Finally, it is only recently dense granular flows have been successfully studied using a continuum model and so, the governing equations have so far been unexplored even for Stokes' first problem. In addition, apart from the simple case of steady Bagnold flow over an inclined plane, the constant shear flow case, the steady state solution of the Stokes' first-problem \citep{cawthorn2011thesis, Andreotti_Forterre_Pouliquen_2011} there exist only few analytical results describing unsteady dense granular flows (see the notable recent exception of \citet{capart2015depth} who gave entrainment rates in the case of transient heap flows from the depth-integrated layer dynamics assuming a local $\mu(I)$--rheology). Therefore, this brief note is aimed at bringing out the key features of this canonical problem as predicted by the $\mu(I)$--rheology. 

The article is set as follows. Firstly, the governing equations are shown to result in a single non-linear shear stress diffusion equation. Its numerical solution is then computed for the case when the wall shear stress is imposed while letting the wall velocity to develop with time. Some approximate unsteady solutions are obtained and compared with computations. Finally, a brief note on the Stokes' first problem with imposed wall speed is given. %However, any comparison between the following results in the context of the $\mu(I)$--rheology and that of a visco-elastic fluid, or for that matter, any other non-Newtonian fluid is beyond the scope of the present work. 

%%%%%%%%%%%%%%%%%%%%%%%%%%%%%%%%%%%%
\section{Governing equations}
\label{sec:equations}
%%%%%%%%%%%%%%%%%%%%%%%%%%%%%%%%%%%%
\subsection{Constitutive laws : $\mu(I)$--rheology}
\label{subsec:MuI}
Analogous to Coulomb's friction law, using dimensional arguments, experiments and numerical simulations, \citet{midi2004dense, iordanoff2004granular, da2005rheophysics} demonstrated that the shear stress $\tau$  is proportional to the normal stress P for $2$D dense granular flows of rigid particles so that $\tau = \mu (I) \mbox{P}$ where the \textit{local} friction coefficient $\mu$ is only a function of a dimensionless parameter called the \textit{Inertial} number $I =\dot{\gamma}d/\sqrt{P/\rho}$. Here, $\dot{\gamma}$ is the local shear rate which is related to the \textit{macroscopic} timescale of the granular flow and $d\sqrt{\rho/\mbox{P}}$ is the \textit{microscopic} timescale corresponding to any local rearrangement of grains of diameter $d$ and density $\rho$ subjected to a local normal stress P. Note that $I$ is also the square-root of the \textit{Savage} or the \textit{Coulomb} number as used in \citet{savage1984mechanics} or \citet{ancey1999theoretical}, respectively. In general, the dimensionless \textit{local} friction coefficient is given by \citep{midi2004dense, jop2006constitutive}
%%%%%%%%%%%%%%%%%%%%%%%%%%%%%%%%%%%
\begin{equation}
\mu (I) = \mu_s + \frac{\mu_d -\mu_s}{\left(1 + \frac{I_0}{I}\right)},
\label{eq:MuI}
\end{equation}
%%%%%%%%%%%%%%%%%%%%%%%%%%%%%%%%%%%
whereby $\mu$ saturates towards two fundamental constants for a dense granular media $\mu_s$ or $\mu_d$ depending respectively on the inertial number $I \ll 1$ (quasi-static regime) or $I \gg 1$ (kinetic or gaseous regime). \citet{jop2006constitutive} proposed a $3$D generalisation of this scalar constitutive relation for a granular material by decomposing the Cauchy stress tensor into an isotropic contribution from the local pressure $p$ and a traceless deviatoric stress tensor $\tau_{ij}$ while assuming that $\tau_{ij}$ is aligned with the strain rate tensor $\dot{\gamma}_{ij} = \frac{1}{2} \left( \partial_i u_j + \partial_j u_i \right)$ (where $u_i$ represent components of the velocity field). So, if $\vec{\textbf{x}}$ is the position vector and $t$ represents time, then Cauchy stress tensor
%%%%%%%%%%%%%%%%%%%%%%%%%%%%%%%%%%%
\begin{equation}
\sigma_{ij}(\vec{\textbf{x}}, t) =  -p(\vec{\textbf{x}}, t) \delta_{ij} + \tau_{ij}(\vec{\textbf{x}}, t),
\label{eq:ConstitutiveLawMuI}
\end{equation}
%%%%%%%%%%%%%%%%%%%%%%%%%%%%%%%%%%%
where $\delta_{ij}$ is the kronecker delta and $\tau_{ij}(\vec{\textbf{x}}, t) =  \eta(\vec{\textbf{x}}, t) \dot{\gamma}_{ij}(\vec{\textbf{x}}, t)$ with the local granular liquid viscosity
%%%%%%%%%%%%%%%%%%%%%%%%%%%%%%%%%%%
\begin{equation}
\eta = \frac{\mu p}{\vert\dot{\gamma}\vert},
\label{eq:ViscosityMuI}
\end{equation}
%%%%%%%%%%%%%%%%%%%%%%%%%%%%%%%%%%%
is, thereby, a non-linear function of the local pressure $p$ and the local second invariant of the strain rate tensor $\vert\dot{\gamma}\vert = \sqrt{\frac{1}{2} \dot{\gamma}_{ij} \dot{\gamma}_{ij}}$ via the local friction coefficient given by \ref{eq:MuI} and the Inertial number for the $3$D case
%%%%%%%%%%%%%%%%%%%%%%%%%%%%%%%%%%%
\begin{equation}
I =\frac{\vert\dot{\gamma}\vert d}{\sqrt{p/\rho}}.
\label{eq:InertialNumberMuI}
\end{equation}
%%%%%%%%%%%%%%%%%%%%%%%%%%%%%%%%%%%
In addition, the solid volume fraction $\phi$ is also a linear function of $I$ \citep[see][p. $238$]{Andreotti_Forterre_Pouliquen_2011} but, in general, it varies very little and so, for the sake of simplicity, it is taken to be a constant in the following.

%%%%%%%%%%%%%%%%%%%%%%%%%%%%%%%%%%%%
\subsection{Stokes' first problem and its steady--state solution}
\label{subsec:StationarySolutions}
Consider an infinite rigid flat plate placed at rest ($t = 0$) on top of a semi-infinite dense granular media. As illustrated in figure \ref{fig:Schematic}, the plate is set to motion impulsively at $t > 0$ by applying a tangential shear stress $\tau_w$ (along $x$-direction) in the presence of a normal stress $p_w$ (along $y$-direction). It is then natural to restrict the analysis to two-dimensions. In fact, the absence of any horizontal length scale implies that the flow properties should depend only on $y$ and $t$. Incompressibility and the initial condition then imply that the vertical velocity is uniformly zero for all $t \geq 0$. And so the only non-zero components of the strain rate tensor are $\dot{\gamma}_{xy} = \dot{\gamma}_{yx} = \frac{1}{2} \left( \partial u / \partial y \right)$, where $u$ is the $x$-component of the velocity field. Therefore, the $2$D shear stress tensor is completely determined by a single scalar shear field $\tau =   \mu p$ so that the $x$ and $y$ momentum equations become, respectively,
%%%%%%%%%%%%%%%%%%%%%%%%%%%%%%%%%%%
\begin{eqnarray}
\phi \rho \frac{\partial u}{\partial t} = -\frac{\partial}{\partial y} \left( \mu p\right),
\label{eq:StokesEqn1}
\end{eqnarray}
%%%%%%%%%%%%%%%%%%%%%%%%%%%%%%%%%%%
\begin{eqnarray}
\frac{\partial p}{\partial y} =  \phi \rho \mbox{g},
\label{eq:StokesEqn2}
\end{eqnarray}
%%%%%%%%%%%%%%%%%%%%%%%%%%%%%%%%%%%
where $\mbox{g}$ is the acceleration due to gravity. Initially the granular media is at rest ($u(y, t = 0) = 0$) but a static granular material can support a wide variety of shear stress and pressure distributions so long as the yield stress in not exceeded so that $\tau \leq \mu_s p$. Within the context of the $\mu(I)$--rheology, the only possible configuration were the granular liquid is at rest corresponds to the case $\tau = \mu_s p$. For all other static states, the constitutive law $\tau = \mu p$ is violated and so, the $\mu(I)$--rheology is no longer applicable. Therefore, in the following, the initial conditions correspond to a specific static state wherein the shear stress equals the yield stress throughout the granular media
%%%%%%%%%%%%%%%%%%%%%%%%%%%%%%%%%%%
\begin{eqnarray}
\tau (y, t = 0) = \mu_s p
\label{eq:InitStokesEqn}
\end{eqnarray}
%%%%%%%%%%%%%%%%%%%%%%%%%%%%%%%%%%%
with $p$ the \textit{hydrostatic} pressure.
Whereas the boundary conditions for all $t > 0$ are
%%%%%%%%%%%%%%%%%%%%%%%%%%%%%%%%%%%
\begin{eqnarray}
p(y = 0, t) &= &p_w  \\
\mu(y = 0, t) & = &\frac{\tau_w}{p_w} = \mu_w ,
\label{eq:BCStokesEqn}
\end{eqnarray}
%%%%%%%%%%%%%%%%%%%%%%%%%%%%%%%%%%%
along with the condition that the grains sufficiently far from the plate remain static i.e. $u(y = \infty, t) = 0$ and so, $\mu = \mu_s$ at $y = \infty$. In contrast to the classical Stokes' problem \citep{stokes1851effect} for a Newtonian fluid, firstly, the frictional force (R.H.S. of eqn. \ref{eq:StokesEqn1}) is not only non-uniform due to the \textit{hydrostatic} pressure $p(y) = p_w + \phi \rho \mbox{g} y$ but also, non-linear since $\mu$ depends on both $p(y)$ and $\partial u / \partial y$ via eqn. \ref{eq:MuI}. And secondly, there exists a non-trivial steady-state solution wherein the shear stress is constant throughout the media such that $\mu p = \tau_w = \mu_w p_w$ as already shown in \citet[pp. $254$-$256$]{Andreotti_Forterre_Pouliquen_2011} and \citet[pp. $46$-$50$]{cawthorn2011thesis}. Since $\mu \in [\mu_s, \mu_d]$ and $p(y)$ increases linearly with the depth $y$, it follows that 
%%%%%%%%%%%%%%%%%%%%%%%%%%%%%%%%%%%
\begin{equation}
\mu = \frac{\mu_w}{1 + \left({\phi \rho \mbox{g} y}/{p_w}\right)},
\label{eq:MuSteadyState}
\end{equation}
%%%%%%%%%%%%%%%%%%%%%%%%%%%%%%%%%%%
for all $y \leq \delta_s$ and $\mu = \mu_s$ otherwise; here the critical depth $\delta_s$ is given by
%%%%%%%%%%%%%%%%%%%%%%%%%%%%%%%%%%%
\begin{equation}
\delta_s = \beta_w \left(\frac{p_w}{\phi \rho \mbox{g}}\right),
\label{eq:ShearLayerThickness}
\end{equation}
%%%%%%%%%%%%%%%%%%%%%%%%%%%%%%%%%%%
with $\beta_w = (\mu_w/\mu_s - 1)$. It denotes the depth beyond which the granular media does not flow $y \geq \delta_s$. Note that the term $\beta_w$ represents the surplus in wall shear stress over the yield criterion at the wall. The steady-state solution \ref{eq:MuSteadyState} can be used to obtain the corresponding velocity profile. As previously shown by \citet[pp. $46$-$50$]{cawthorn2011thesis}, the resulting relation between the wall velocity (if no-slip condition is allowed) and the shear layer thickness compares qualitatively well with the molecular dynamic simulations of \citet{Thompson_PRL1991granular}. 

%%%%%%%%%%%%%%%%%%%%%%%%%%%%%%%%%%%%
\subsection{Stokes' first problem: shear stress diffusion equation}
\label{subsec:GoverningEqnShearStress}
Most numerical studies obtain the velocity field by solving the momentum equations that account for the constitutive law \ref{eq:ConstitutiveLawMuI}. Note that the latter is coupled with the expressions for the local friction coefficient (\ref{eq:MuI}) and the inertial number $I$. However, it is possible to write a single equation for the shear stress in the case of the Stokes' first problem with in the context of the $\mu(I)$--rheology. Since $I=-\frac{1}{2} \frac{\partial u}{\partial y} \sqrt{\frac{\rho d^2}{p}}$, the local friction coefficient \ref{eq:MuI} can be rewritten as
%%%%%%%%%%%%%%%%%%%%%%%%%%%%%%%%%%%
\begin{equation}
\frac{\partial u}{\partial y} = - 2 I_0 \sqrt{\frac{p}{\rho d^2}} \left( \frac{\mu - \mu_s}{\mu_d - \mu} \right),
\label{eq:IreWritten}
\end{equation}
%%%%%%%%%%%%%%%%%%%%%%%%%%%%%%%%%%%
which when differentiated w.r.t. time and using $\tau = \mu p$ yields
\begin{equation}
\frac{\partial^2 u}{\partial t \partial y} = - \frac{2 \Delta \mu I_0}{d} \left(p \sqrt{\frac{p}{\rho}} \right) \frac{1}{\left( \mu_d p - \tau \right)^2}\frac{\partial \tau}{\partial t},
\label{eq:step2StokesEqn}
\end{equation}
%%%%%%%%%%%%%%%%%%%%%%%%%%%%%%%%%%%
where $\Delta \mu = \mu_d - \mu_s > 0$. By allowing $\frac{\partial^2 u}{\partial y \partial t} = \frac{\partial^2 u}{\partial t \partial y}$ and then introducing \ref{eq:StokesEqn1} in the above expression, it leads to a single non-linear shear-stress diffusion equation
%%%%%%%%%%%%%%%%%%%%%%%%%%%%%%%%%%%
\begin{equation}
\frac{\partial \tau}{\partial {t}} = \left( \frac{d}{2 \phi \Delta \mu I_0 \sqrt{\rho}} \right)\frac{\left(\mu_d {p} - {\tau} \right)^2}{p \sqrt{p}} \frac{\partial^2 \tau}{\partial {y}^2}.
\label{eq:GranularStokesEqnDIM}
\end{equation}
%%%%%%%%%%%%%%%%%%%%%%%%%%%%%%%%%%%
Finally, by taking $\nu_g = (d/2\phi \Delta \mu I_0) \sqrt{p_w/\rho}$ as a proper diffusion coefficient and the steady--state shear layer thickness $\delta_s$ as the characteristic length scale, the non-dimensional time and space coordinates are  $\tilde{t} = {\nu_g t}/{\delta_s^2}$ and $\tilde{y} = {y}/{\delta_s}$, respectively. Thus, in terms of the normalised pressure $\tilde{p} = {p(y)}/{p_w}$ and shear stress $\tilde{\tau} = {\tau(y, t)}/{\tau_w}$, the above equation becomes
%%%%%%%%%%%%%%%%%%%%%%%%%%%%%%%%%%%
\begin{equation}
%\frac{\partial \bar{\mu}}{\partial \bar{t}} = \frac{\left( \bar{\mu} + \mu_d \bar{y} - \mu_w \right)^2}{\bar{y}\sqrt{\bar{y}}} \frac{\partial^2 \bar{\mu}}{\partial \bar{y}^2},
\frac{\partial \tilde{\tau}}{\partial \tilde{t}} = \frac{\left(\mu_d \tilde{p} - \mu_w \tilde{\tau} \right)^2}{\tilde{p} \sqrt{\tilde{p}}} \frac{\partial^2 \tilde{\tau}}{\partial \tilde{y}^2},
%\frac{\partial \tilde{\tau}}{\partial {t}} = \nu_g \frac{\left(\mu_d \tilde{p} - \mu_w \tilde{\tau} \right)^2}{\tilde{p} \sqrt{\tilde{p}}} \frac{\partial^2 \tilde{\tau}}{\partial {y}^2},
\label{eq:GranularStokesEqn}
\end{equation}
%%%%%%%%%%%%%%%%%%%%%%%%%%%%%%%%%%%
with boundary conditions $\tilde{\tau}(0, \tilde{t}) = \tilde{\tau}(1, \tilde{t}) = 1$ and an initial condition $\tilde{\tau}(\tilde{y}, 0) = \mu_s \tilde{p}/\mu_w$. The steady-state solution for the non-dimensional shear stress is $\tilde{\tau} = 1$. Unlike the steady-state local friction coefficient $\mu$, the steady-state shear stress is a continuous and infinitely differentiable function for all $y \geq 0$. So, it is expected that $\tilde{\tau}$ remains smooth also for the unsteady case. In fact, the term $(\mu_d \tilde{p} - \mu_w \tilde{\tau})$ is positive--definite. Hence, it is quite straight-forward to homogenise the boundary conditions and numerically solve the above equation using a second-order centred finite difference scheme for spatial derivatives and a second-order Crank-Nicholson one for temporal integration. The updated $\tilde{\tau}$ is obtained by an iterative Richardson Minimal Residual process.
%%%%%%%%%%%%%%%%%%%%%%%%%%%%%%%%%%%%
\subsection{Well-posedness of the shear stress diffusion equation}
\label{subsec:GoverningEqnShearStressHadamard}
\citet{barker2015well} demonstrated that the $\mu(I)$--rheology is well-posed for intermediate values of inertial number $I$, but that
it is ill-posed for both high and low inertial numbers. In the present case, $I \gg 1$ in the neighbourhood of the wall either when the applied wall-shear $\tau_w$ is close to the critical wall shear $\tau_d = \mu_d p_w$ or when wall speed $u_w$ is sufficiently large. In addition, there is always a zone where $I \ll 1$ (or $\mu \sim \mu_s)$ since the media is slowly-moving or stationary when $y \sim \delta_s$. As already shown by \citet{barker2015well} this should provoke \textit{Hadamard instability} \citep{joseph1990short} whereby infinitesimally small short-wave perturbations are amplified indefinitely. Thus, numerical solutions may not converge as the grid is refined and so, before proceeding any further, it is important to verify that the shear stress diffusion equation \ref{eq:GranularStokesEqn} is indeed well-posed for all values of inertial number $I$. 

%According to \citet{joseph1990short}, such ill-posedness in governing equations can be identified with \textit{Hadamard instability} wherein infinitesimally small short-wave perturbations are amplified indefinitely. Note that
%It is not necessary to precisely know the solution to the governing equation nor to follow a complete linear stability analysis. Instead, it is necessary to retain only the highest order derivatives of each perturbed field with respect to each spatial coordinate since \textit{Hadamard instability} is related to high-wavenumber limit of the perturbed field \citep{barker2015well, martin2017continuum}.

In fact, this non-linear diffusion equation \ref{eq:GranularStokesEqn} is one-dimensional and when it is linearized about an arbitrary base state as in \citet[see pp. 799]{barker2015well}, the resulting dispersion equation of the normal mode analysis in the high-wavenumber limit should be
\begin{equation}
	\lambda = - \alpha^2 \xi_y^2,
\label{eq:DispersionEquationHadamard}
\end{equation}
where $\lambda$ is the complex frequency, $\xi_y$ is the wave number in the $\tilde{y}$-direction and  $\alpha = \alpha(\tilde{p}^{0}, \tilde{\tau}^{0})$ is some function of the base state pressure $\tilde{p}^{0}$ and shear stress $\tilde{\tau}^{0}$. Since the real-part of the complex frequency $\lambda$ provides the perturbation growth rate, it is straight-forward to see that the equation \ref{eq:GranularStokesEqn} is stable for short-waves along the $y$-coordinate.

Note that this is an one-dimensional analysis since it takes into account only plane waves along the $\tilde{y}$-axis in order to analyse the well-posedness of \ref{eq:GranularStokesEqn} for one-dimensional, time dependent computations. For two-dimensional codes that consider granular Stokes problem as a test case, the situation remains more complex. In this case, as previously shown by \citet{barker2015well}, the equations are still ill-posed since oblique two-dimensional short waves are unstable in the region $y \sim \delta_s$ (or close to the wall for $\mu_w \sim \mu_d$).
%%%%%%%%%%%%%%%%%%%%%%%%%%%%%%%%%%%%
\section{Unsteady solutions and their characteristics}
\label{sec:UnsteadySolution}
\begin{figure}
\begin{center}
\epsfig{file=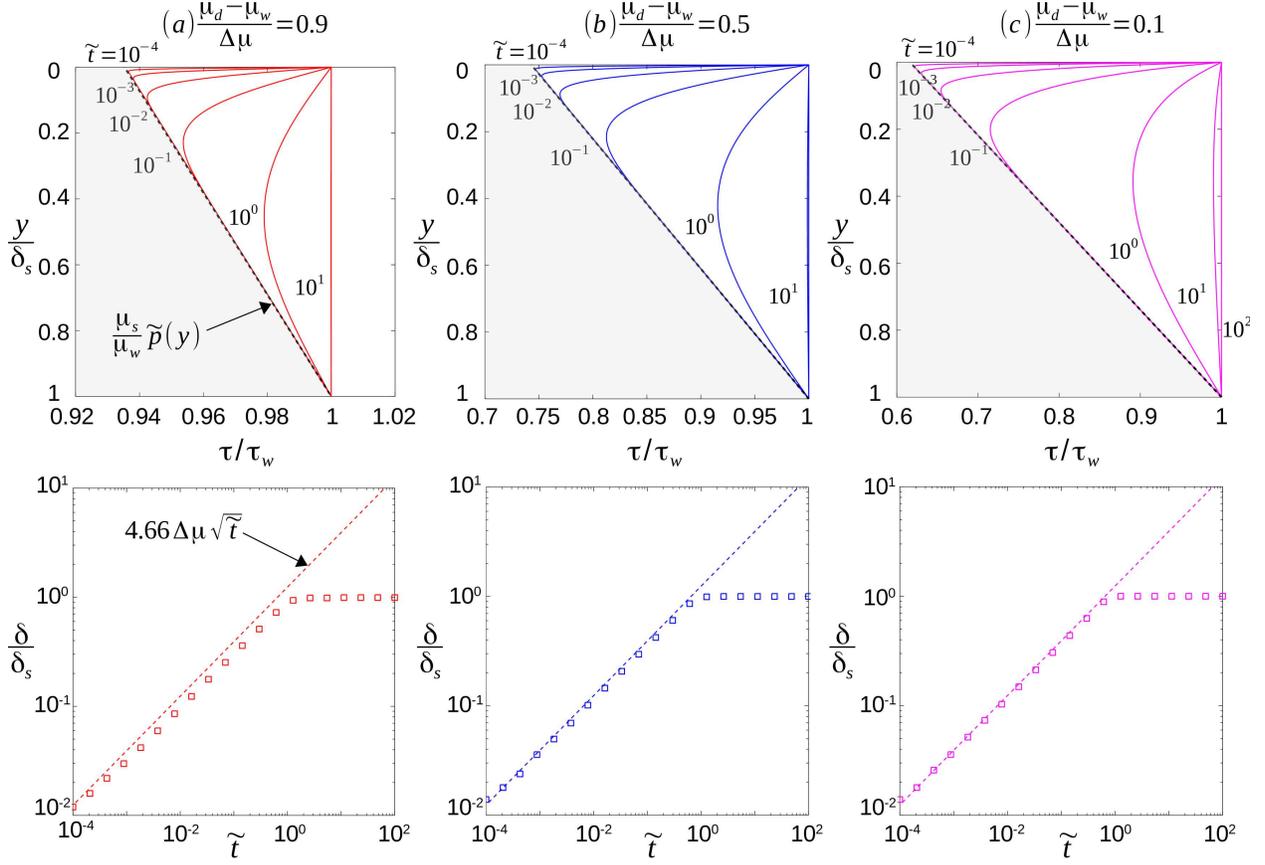,width=1\textwidth,keepaspectratio=true}
\end{center}
\caption{Temporal evolution of the normalised shear stress $\tilde{\tau} = \tau/\tau_w = \mu \tilde{p}/\mu_w$ (top) and  shear layer thickness $\delta/ \delta_s$ (bottom) for typical values of the normalised wall friction coefficient $(\mu_d - \mu_w)/\Delta \mu$ (a) $0.9$, (b) $0.5$ \& (c) $0.1$ as computed directly using \ref{eq:GranularStokesEqn} with initial condition $\tilde{\tau}(0, \tilde{y})= \mu_s \tilde{p}/\mu_w$ corresponding to a no flow regime (shaded region). Note that $\Delta \mu = \mu_d - \mu_s$.}% In all cases, the thickness of the region where the shear stress is more than the yield shear stress $\mu_s p$ grows as $\sqrt{\tilde{t}}$ until very rapidly saturating at $\delta_s = (\mu_w / \mu_s -1)(p_w/\phi \rho g)$ when $\tilde{t}$ is O$(1)$.}
\label{fig:Fig1_ShearStressNum}
\end{figure}

Figure \ref{fig:Fig1_ShearStressNum} presents the results of such numerical solutions (continuous lines) for three typical values of applied wall shear stress when $\mu_s = \tan 21^{\circ}$ and $\mu_d = \tan 33^{\circ}$ (typical values for spherical mono--dispersed glass beads as in \citet{Andreotti_Forterre_Pouliquen_2011}). Each graphic (top) depicts the normalised shear stress $\tilde{\tau} = \mu \tilde{p}/\mu_w$ variation in the $y$-direction at various time $\tilde{t} = 10^{-4}, 10^{-3}, 10^{-2}, 10^{-1}, 10^{0}, 10^{1}$ for typical values of the normalised wall friction coefficient $(\mu_d - \mu_w)/\Delta \mu$ (where, $\Delta \mu = \mu_d - \mu_s$). In all cases, the static initial condition $\tilde{\tau}(0, \tilde{y}) = \mu_s \tilde{p}/\mu_w$ (dashed line), wherein the local shear stress is taken to be the yield criterion $\mu_s p$, evolves continuously towards the steady--state solution $\tilde{\tau} = \mu \tilde{p}/\mu_w = 1$. The spatial variation of $\tilde{\tau}$ shows that there exists a layer in which $\tilde{\tau}$ is greater than the yield stress and so, the granular media should flow in this region. If the size of this shear layer, say $\delta(\tilde{t})$, is defined as the region where $\tilde{\tau} = 0.999 \mu_s \tilde{p}/\mu_w$, figure \ref{fig:Fig1_ShearStressNum} (bottom) clearly illustrates that $\delta(\tilde{t})$ increases with time as $\sqrt{\tilde{t}}$ until $\tilde{t} \sim \mbox{  \textit{O}} (1)$, after which it saturates to the steady-state limit. Therefore, it is expected from these results that, for any general $\mu_d$ and $\mu_s$, approximate solutions to \ref{eq:GranularStokesEqn} can be obtained at both $\tilde{t} \ll 1$ and $\tilde{t} \gg 1$ by properly linearising it.

%%%%%%%%%%%%%%%%%%%%%%%%%%%%%%%%%%%%
\subsection{Self-similarity at $\tilde{t} \ll 1$}
\label{subsec:SelfSimilarSolution}
When $\tilde{t} \ll 1$, the non-dimensional shear layer thickness $\delta/\delta_s$ is small as observed in figure \ref{fig:Fig1_ShearStressNum} (bottom). By taking the non-dimensional local pressure $\tilde{p} = 1 + \beta_w \tilde{y}$ and local shear stress $\tilde{\tau} = \mu (1 + \beta_w \tilde{y})/\mu_w$, the diffusion equation \ref{eq:GranularStokesEqn} becomes
%%%%%%%%%%%%%%%%%%%%%%%%%%%%%%%%%%%%
\begin{equation}
\frac{\partial \mu}{\partial \tilde{t}} = \left(\mu_d - \mu \right)^2 \left( \frac{2 \beta_w}{\sqrt{1 + \beta_w \tilde{y}}} \frac{\partial \mu}{\partial \tilde{y}} + \sqrt{1 + \beta_w \tilde{y}} \frac{\partial^2 \mu}{\partial \tilde{y}^2} \right),
\label{eq:GranularStokesEqn_mu}
\end{equation}
%%%%%%%%%%%%%%%%%%%%%%%%%%%%%%%%%%%%
with $\mu(\tilde{y}, 0) = \mu_s$, $\mu(0, \tilde{t}) = \mu_w$ and $\mu(1, \tilde{t}) = \mu_s$. Note that, in general, $\beta_w < 1$ and hence, for $\tilde{t} \ll 1$, the highest--order derivative of $\mu$ should dominate if $\tilde{y} \leq \delta/\delta_s$ so that $\mu \sim \mu_w - (\mu_w - \mu_s) y/\delta$. And in the outer region, $\mu \sim \mu_s$. As the spatial variations of the local friction coefficient $\mu$ are stronger inside the shear layer (i.e., when $\beta_w \tilde{y} \ll 1$), it is reasonable to simplify \ref{eq:GranularStokesEqn_mu} to
%%%%%%%%%%%%%%%%%%%%%%%%%%%%%%%%%%%%%
\begin{equation}
\frac{\partial \mu}{\partial \tilde{t}} = \Delta \mu^2  \frac{\partial^2 \mu}{\partial \tilde{y}^2},
\label{eq:GranularStokesEqn_muLINshort}
\end{equation}
%%%%%%%%%%%%%%%%%%%%%%%%%%%%%%%%%%%%%
at the leading order with the same boundary conditions as before. By taking $\psi = \mu - \left( \mu_w - (\mu_w - \mu_s)\tilde{y} \right)$, the above equation admits a self-similar solution for $\psi = \psi(\eta)$ with $\eta = \tilde{y}/(2 \Delta \mu \sqrt{\tilde{t}})$ satisfying the initial and the boundary conditions for all $\tilde{t} \geq 0$. Thereby, the local friction coefficient is deduced to be
%%%%%%%%%%%%%%%%%%%%%%%%%%%%%%%%%%%%%
\begin{equation}
\mu \sim \mu_s + (\mu_w - \mu_s)(1 - \tilde{y}) \mbox{ erfc } \left( \frac{\tilde{y}}{2 \Delta \mu \sqrt{\tilde{t}}} \right),
\label{eq:SolutionSS}
\end{equation}
%%%%%%%%%%%%%%%%%%%%%%%%%%%%%%%%%%%%%
which implies that the shear layer thickness should grow as $\delta / \delta_s \sim 4.66 \Delta \mu \sqrt{\tilde{t}}$. 

The expression \ref{eq:SolutionSS} is presented in figure \ref{fig:Fig2_ShearStressComparison} (top) where the time evolution of the local friction coefficient is displayed as function of $\tilde{y}$ at different time $\tilde{t} = 10^{-4}, 10^{-3}, 10^{-2}, 10^{-1}, 10^{0}, 10^{1}$ for the same values of $(\mu_d - \mu_w)/\Delta \mu$ as in figure \ref{fig:Fig1_ShearStressNum}. When compared with numerical solutions of $\mu$ (and also, $\delta$) as seen in figure \ref{fig:Fig2_ShearStressComparison} (top) (and figure \ref{fig:Fig1_ShearStressNum}, respectively), these approximations are very satisfactory for all $\tilde{t} \ll 1$. Indeed, the expression \ref{eq:SolutionSS} for the local friction coefficient and especially, the estimations of the time evolution of the shear layer thickness $\delta / \delta_s \sim 4.66 \Delta \mu \sqrt{\tilde{t}}$ are reasonably good even when $\tilde{t}$ is of order $1$.
%%%%%%%%%%%%%%%%%%%%%%%%%%%%%%%%%%%%%

%%%%%%%%%%%%%%%%%%%%%%%%%%%%%%%%%%%%%
\subsection{Diffusion at $\tilde{t} \gg 1$}
\label{subsec:LongTimeSolution}
%%%%%%%%%%%%%%%%%%%%%%%%%%%%%%%%%%%
\begin{figure}
\begin{center}
\epsfig{file=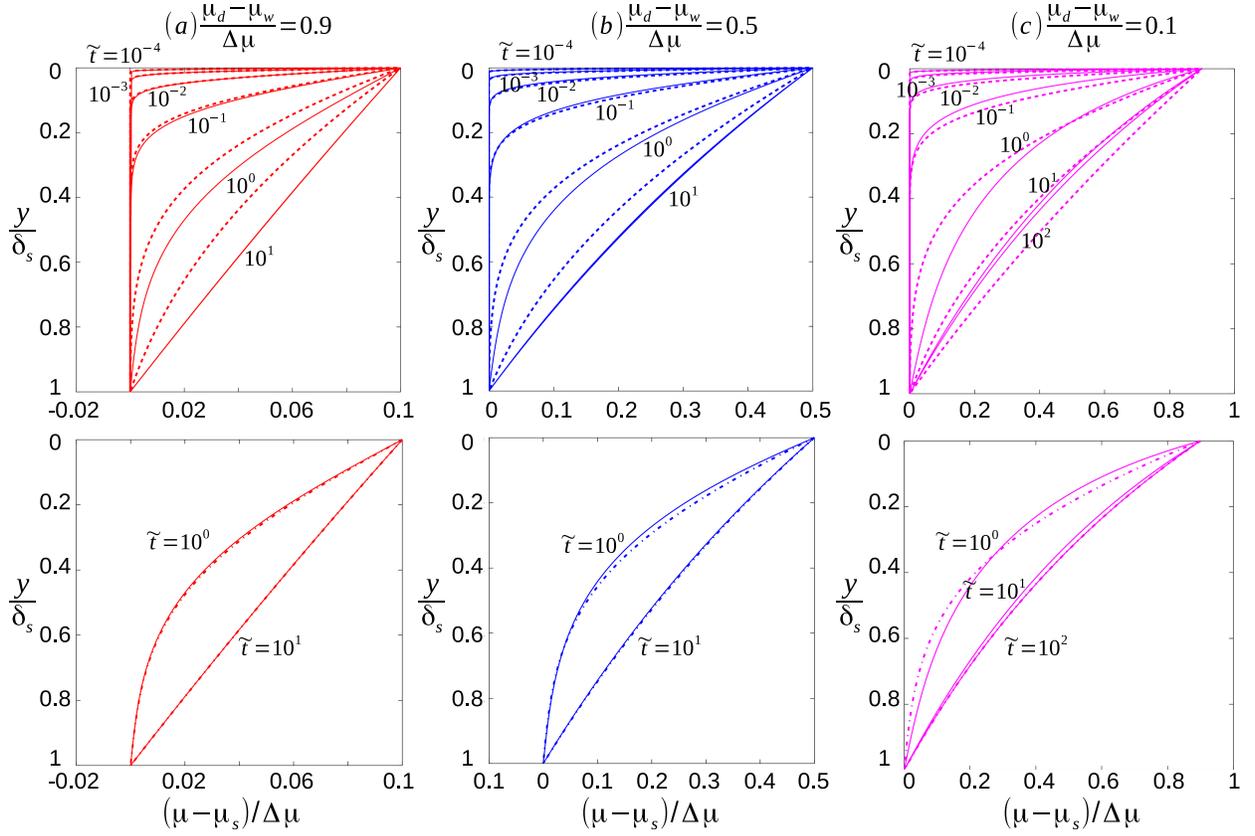,width=1\textwidth,keepaspectratio=true}
\end{center}
\caption{Comparison between self-similar approximation (top) and the long-time diffusion approximation (bottom) with the direct numerical solution (continuous lines) for various normalised wall friction coefficients $(\mu_d - \mu_w)/(\mu_d - \mu_s)$ (a) $0.9$, (b) $0.5$ \& (c) $0.1$.}
\label{fig:Fig2_ShearStressComparison}
\end{figure}
%%%%%%%%%%%%%%%%%%%%%%%%%%%%%%%%%%%%
As soon as $\tilde{t} \sim \mbox{  \textit{O}} (1)$ the non-dimensional shear layer thickness is no longer small and hence, a singular perturbation of \ref{eq:GranularStokesEqn_mu} cannot be obtained with the present scaling for the $y$-coordinate. However, by using the non-dimensional pressure $\tilde{p}$ as an equivalent normalised spatial variable $\hat{y} = 1 + \beta_w \tilde{y}$, it is possible to show that
%%%%%%%%%%%%%%%%%%%%%%%%%%%%%%%%%%%%%
\begin{equation}
\frac{\partial \tilde{\tau}}{\partial \tilde{t}} = (\beta_w (\mu_d - \mu))^2 \sqrt{\hat{y}} \frac{\partial^2 \tilde{\tau}}{\partial \hat{y}^2},
\label{eq:GranularStokesEqn_muLINlong}
\end{equation}
%%%%%%%%%%%%%%%%%%%%%%%%%%%%%%%%%%%%%
which is singular if $(\beta_w \Delta \mu)^2$ tends to zero. This is often true since $(\beta_w (\mu_d - \mu))^2$ is of the order of $(\beta_w \Delta \mu)^2$ and $\hat{y} \sim \mbox{  \textit{O}} (1)$. Thus, taking $\beta_w (\mu_d - \mu) \sim \beta_w \Delta \mu$ at the leading order in \ref{eq:GranularStokesEqn_muLINlong}, it becomes linear and admits a WKB--approximation in the $\hat{y}$-coordinate. Thereby, the local shear stress $\tilde{\tau}$ can be shown to be
%%%%%%%%%%%%%%%%%%%%%%%%%%%%%%%%%%%%%
\begin{equation}
\tilde{\tau} \sim 1 + \hat{y}^{-1/8} \sum^{\infty}_{m = 1} \left[ \Lambda_m \exp{\left(- \lambda_m^2 \tilde{t}\right)} \sin{\left( m \pi \frac{\hat{y}^{3/4} - 1}{(\mu_w/\mu_s)^{3/4} - 1} \right)}\right],
\label{eq:SolutionLONG}
\end{equation}
%%%%%%%%%%%%%%%%%%%%%%%%%%%%%%%%%%%%%
where
%%%%%%%%%%%%%%%%%%%%%%%%%%%%%%%%%%%%%
\begin{eqnarray}
\Lambda_m &= &2 \int^{\mu_w/\mu_s}_{1} {\hat{y}^{1/8} \left( \frac{\mu_s}{\mu_w} \hat{y} - 1 \right) \sin{\left( m \pi \frac{\hat{y}^{3/4} - 1}{(\mu_w/\mu_s)^{3/4} - 1} \right)} } d \hat{y}, \\
\lambda_m &= &\frac{3}{4} \frac{m \pi \beta_w \Delta \mu}{(\mu_w/\mu_s)^{3/4} - 1}.
\label{eq:SolutionLONG_constants}
\end{eqnarray}
%%%%%%%%%%%%%%%%%%%%%%%%%%%%%%%%%%%%%
Note that it is possible to solve the linearised version of \ref{eq:GranularStokesEqn_muLINlong} directly by separation of variables as well. In that case, the general solution is an infinite sum of Bessel functions in the $\hat{y}$--direction. But it is advantageous to work with the approximate solution \ref{eq:SolutionLONG} when $m$ is large. Nevertheless, the comparison between the WKB--approximation \ref{eq:SolutionLONG} (dot-dash lines in figure \ref{fig:Fig2_ShearStressComparison}) and the numerical solution for the local friction coefficient is good. And, in particular, the agreement is excellent when the applied friction coefficient $\mu_w$ is close to yield friction coefficient $\mu_s$.
%\section{Discussion}
%\label{sec:Discussion}
%%%%%%%%%%%%%%%%%%%%%%%%%%%%%%%%%%%%
\subsection{Velocity field development}
\label{subsec:BL}
It is now possible to compute the temporal evolution of the velocity field using local shear stress $\tau = \mu p$. By rewriting eqn. \ref{eq:StokesEqn1} in terms of $\tilde{t}$ and $\tilde{y}$, a natural normalisation for the velocity $u$ can be shown to be
%as $u = \tilde{u} \left({g \delta_s^2}/{\beta_w \nu_g}\right)$
%%%%%%%%%%%%%%%%%%%%%%%%%%%%%%%%%%% 
\begin{equation}
	u = \tilde{u} \left(\frac{g \delta_s^2}{\beta_w \nu_g}\right),
	\label{eq:nonDIMv}
\end{equation}
%%%%%%%%%%%%%%%%%%%%%%%%%%%%%%%%%%% 
where $\delta_s$ is the steady-state shear layer thickness (eqn. \ref{eq:ShearLayerThickness}), $\beta_w = (\mu_w - \mu_s)/\mu_s$ and $\nu_g  = (d/\phi \Delta \mu I_0) \sqrt{p_w/\rho}$ is the diffusion coefficient which appears in the granular Stokes' equation \ref{eq:GranularStokesEqnDIM}. Using this normalisation and by taking 
%%%%%%%%%%%%%%%%%%%%%%%%%%%%%%%%%%% 
\begin{equation}
	\tilde{\mu} = \frac{\mu_d - \mu}{\Delta \mu},
	\label{eq:nonDIMmu}
\end{equation}
%%%%%%%%%%%%%%%%%%%%%%%%%%%%%%%%%%% 
the equation for the shear rate \ref{eq:IreWritten} becomes
%%%%%%%%%%%%%%%%%%%%%%%%%%%%%%%%%%%
\begin{equation}
\frac{\partial \tilde{u}}{\partial \tilde{y}} = -\frac{\sqrt{\tilde{p}}}{\Delta \mu} \left( \frac{1}{\tilde{\mu}} - 1 \right),
\label{eq:IreWrittenNonDim}
\end{equation}
%%%%%%%%%%%%%%%%%%%%%%%%%%%%%%%%%%% 
which can then be integrated to study the velocity field. In the following, no-slip condition is assumed so that the wall speed and the velocity of granular media right next to it ($\tilde{y} = 0$) are equal.
\begin{figure}
\begin{center}
\epsfig{file=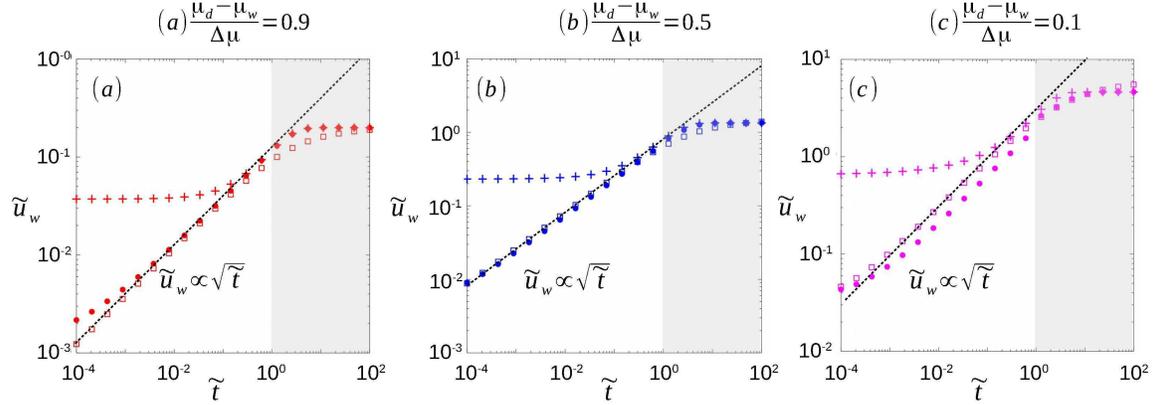,width=1\textwidth,keepaspectratio=true}
\end{center}
\caption{Normalised wall velocity as a function of the non-dimensional time $\tilde{t}$ for various applied wall shear stress : $(\mu_d - \mu_w)/(\mu_d - \mu_s)$ (a) $0.9$, (b) $0.5$ \& (c) $0.1$. The different symbols denote calculations from numerical solution using \ref{eq:GranularStokesEqn} (filled circles), self-similar solution \ref{eq:SolutionSS} ($\square$) and the long--time approximate solution \ref{eq:SolutionLONG} ($+$).}% For all values of $(\mu_d - \mu_w)/(\mu_d - \mu_s)$, the wall velocity increases as $\sqrt{\tilde{t}}$ (dashed line) in good comparison with self-similar solution until about $\tilde{t} \sim \mbox{ O}(1)$.}% It then saturates to the steady--state value following the WKB approximation \ref{eq:SolutionLONG} when $\tilde{t} \geq 1$.}
\label{fig:Fig3_WallSpeedComparison}
\end{figure}

Figure \ref{fig:Fig3_WallSpeedComparison} compares normalised wall velocity $\tilde{u}_w = \tilde{u}(\tilde{y} = 0, \tilde{t})$ when obtained from the numerical solution of \ref{eq:GranularStokesEqn} (filled circles), self-similar solution \ref{eq:SolutionSS} ($\square$) and the long--time approximate solution \ref{eq:SolutionLONG} ($+$) at a given boundary condition at the wall. Each graphic shows the temporal evolution of $\tilde{u}_w$ when the wall friction coefficient is taken as $(\mu_d - \mu_w)/(\mu_d - \mu_s)$ (a) $0.9$, (b) $0.5$ \& (c) $0.1$. In all cases, the wall velocity grows monotonically by following a power law in $\tilde{t}$ as long as $\tilde{t} \leq 1$ and then it saturates at the steady state value $\tilde{u}_w^{\infty} = \tilde{u}(\tilde{y} = 0, \tilde{t} = \infty)$. Firstly, the self-similar solution \ref{eq:SolutionSS} gives a good qualitative agreement with numerical results and also, it matches well with the power law $\tilde{u}_w(\tilde{t}) \sim \sqrt{\tilde{t}}$. Whereas the long--time WKB approximation \ref{eq:SolutionLONG} matches very well with the numerical solution at all times $\tilde{t} > 1$. Secondly, it is observed that the steady-state wall velocity $\tilde{u}_w^{\infty}$ depends largely on the imposed wall-shear via $\tilde{\mu}_w = \left( \mu_d - \mu_w \right)/\Delta \mu$ as already shown by previous works \citep{Andreotti_Forterre_Pouliquen_2011, cawthorn2011thesis}. In fact, for the steady state solution, \citet[pp.49]{cawthorn2011thesis} had already given an asymptotic solution for the wall speed  $\tilde{u}_w^{\infty}$ when the applied shear stress is just above the yield shear i.e, when $\mu(\tilde{y} = 0, \tilde{t})$ approaches $\mu_s$ from above (or $\tilde{\mu}_w \sim 1$). However, it is also possible to obtain expressions for $\tilde{u}_w^{\infty}$ for a wide range of $\tilde{\mu}_w$  via the self-similar solution \ref{eq:SolutionSS} since, as suggested by figure \ref{fig:Fig3_WallSpeedComparison}, it gives a good approximation to the steady state wall speed.  In the limit when $\tilde{t} \gg 1$, the approximate solution \ref{eq:SolutionSS} becomes a function only of $\tilde{y}$ and so, in terms of the normalised friction coefficient \ref{eq:nonDIMmu}, it is given by $\tilde{\mu} \sim \tilde{\mu}_w + (1 - \tilde{\mu}_w)\tilde{y}$. Using this expression in \ref{eq:IreWrittenNonDim}, it reads
%%%%%%%%%%%%%%%%%%%%%%%%%%%%%%%%%%%
\begin{equation}
\tilde{u}_w^{\infty} \sim \int_{0}^{1} { \frac{\sqrt{1 + \beta_w \tilde{y}}}{\Delta \mu} \left( \frac{1}{\tilde{\mu}_w + (1 - \tilde{\mu}_w)\tilde{y}} -1 \right)} d{\tilde{y}},
\label{eq:UwallInt}
\end{equation}
%%%%%%%%%%%%%%%%%%%%%%%%%%%%%%%%%%%
and since $\beta_w = \left(\mu_w/\mu_s -1 \right) < 1$, it could be further developed to obtain
%%%%%%%%%%%%%%%%%%%%%%%%%%%%%%%%%%%%
%\begin{equation}
%\tilde{u}_w^{\infty} \sim \int_{0}^{1} { \frac{1}{\Delta \mu} \left( \frac{1}{\tilde{\mu}_w + (1 - \tilde{\mu}_w)\tilde{y}} -1 \right)} d{\tilde{y}} + \mathcal{O}(\beta_w),
%\label{eq:UwallIntbis}
%\end{equation}
%%%%%%%%%%%%%%%%%%%%%%%%%%%%%%%%%%%%
%which finally yields
a simple expression for the the steady-state wall velocity
%%%%%%%%%%%%%%%%%%%%%%%%%%%%%%%%%%%%%%%%
\begin{equation}
\tilde{u}_w^{\infty} \sim -\frac{1}{\Delta \mu} \left[ 1 + \frac{\log{\tilde{\mu}_w}}{(1 - \tilde{\mu}_w)} \right]+ \mathcal{O}(\beta_w).
\label{eq:Uwall_asymp}
\end{equation}
%%%%%%%%%%%%%%%%%%%%%%%%%%%%%%%%%%%%%%%%
Noting that $\tilde{\mu}_w = 1 - \mu_s \beta_w/ \Delta \mu$, it is straightforward to see that the first term in the above expression cancels out when $\tilde{\mu}_w \sim 1$ (or $\beta_w = (\mu_w/\mu_s -1) \ll 1$) and thereby, it gives $\tilde{u}_w^{\infty} \sim \mathcal{O}(\beta_w)$. In this case, as already deduced by \citet[pp.49]{cawthorn2011thesis}, the above integral leads to
%%%%%%%%%%%%%%%%%%%%%%%%%%%%%%%%%%%%%%%%
%%%and since $\tilde{\mu}_w = 1 - \mu_s \beta_w/ \Delta \mu$, the above integral leads to
%%%%%%%%%%%%%%%%%%%%%%%%%%%%%%%%%%%%%%
%%%\begin{equation}
%%%\tilde{u}_w^{\infty} \sim \int_{0}^{1} { \frac{\sqrt{1 + \beta_w \tilde{y}}}{\Delta \mu} \left( \frac{\mu_s \beta_w}{\Delta \mu} (1 - \tilde{y}) \right)} d{\tilde{y}},
%%%\label{eq:UwallInt_muS}
%%%\end{equation}
%%%%%%%%%%%%%%%%%%%%%%%%%%%%%%%%%%%%%%%%
%%%when $\tilde{\mu}_w \sim 1$ (or $\beta_w = (\mu_w/\mu_s -1) \ll 0$). Thus for this case, the steady-state wall velocity is
%%%%%%%%%%%%%%%%%%%%%%%%%%%%%%%%%%%%%
\begin{equation}
\tilde{u}_w^{\infty} \sim \frac{\mu_s}{2 \Delta \mu^2} \left( \beta_w - \frac{5}{2} \beta_w^2 + \mathcal{O}(\beta_w^3) \right).
\label{eq:Uwall_muS}
\end{equation}
%%%%%%%%%%%%%%%%%%%%%%%%%%%%%%%%%%%%%
For the case when $\tilde{\mu}_w$ tends to zero (or $\mu \rightarrow \mu_d$), the integral \ref{eq:UwallInt} will exhibit a logarithmic singularity at $\tilde{y} = 0$ and so, $\tilde{u}_w^{\infty} \sim -{\log{\tilde{\mu}_w}}/{\Delta \mu}$. For the sake of completeness, it is pointed out that, even when $\beta_w$ is not smaller than one, an expression similar to \ref{eq:Uwall_asymp} could be developed at $\tilde{\mu}_w \ll 1$ by exploiting the logarithmic singularity in the integral \ref{eq:UwallInt}.
\begin{figure}
\begin{center}
\epsfig{file=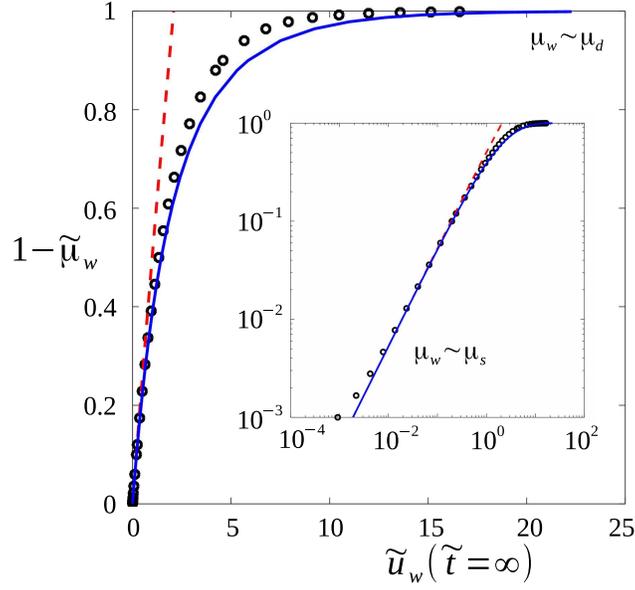,width=0.5\textwidth,keepaspectratio=true}
\end{center}
\caption{Comparison between direct computations and approximate expressions of the steady-state wall speed $\tilde{u}_w^{\infty} = \tilde{u}(\tilde{y} = 0, \tilde{t} \gg 1)$ as a function for the entire range of normalised wall friction coefficient $\tilde{\mu}_w = (\mu_d - \mu_w)/(\mu_d - \mu_s)$. Black circles denote the numerical solution using \ref{eq:GranularStokesEqn}. The continuous line (blue) and the dashed line (red) are obtained using the expressions \ref{eq:Uwall_asymp} \& \ref{eq:Uwall_muS}, respectively. The latter gives a good match when the applied shear stress is just above the yield shear stress $\tilde{\mu}_w \sim 1$ (or $\mu_w \sim \mu_s$). Whereas the former captures the trend for all values of  $\tilde{\mu}_w \in [0, 1] $.}
\label{fig:Fig4_WallSpeedSS}
\end{figure}

\begin{figure}
\begin{center}
\epsfig{file=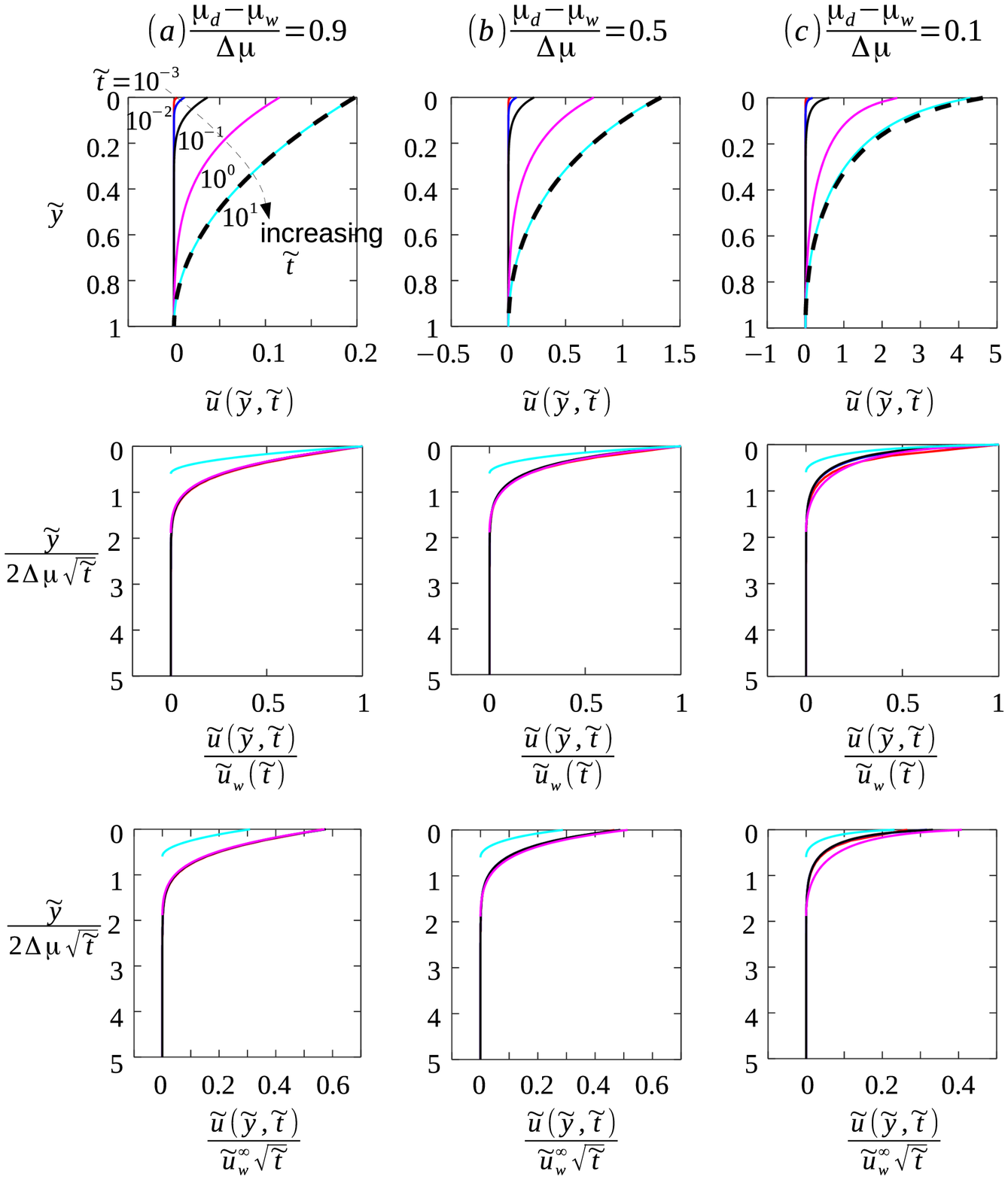,width=1\textwidth,keepaspectratio=true}
\end{center}
\caption{Velocity profile at different time $\tilde{t} = 10^{-3}, 10^{-2}, 10^{-1}, 10^{0}, 10^{1}$ in the shear layer for various normalised wall-friction coefficient $\tilde{\mu}_w = 0.9$ (left),  $\tilde{\mu}_w = 0.5$ (middle) and  $\tilde{\mu}_w = 0.1$ (right). Thick dashed line represents the steady-state solution. These data are obtained by directly integrating the numerical solution of the non-dimensional shear stress equation \ref{eq:GranularStokesEqn}. The data collapse in the second and third rows implies that the $\tilde{u}(\tilde{y}, \tilde{t}) \sim \tilde{u}_w^{\infty}\sqrt{\tilde{t}} f\left( \frac{\tilde{y}}{2 \Delta \mu \sqrt{\tilde{t}}} \right)$, where $\tilde{u}_w^{\infty}$ is given by the expression \ref{eq:Uwall_asymp}.}
\label{fig:Fig5_VelocityProfile}
\end{figure}
The expressions \ref{eq:Uwall_asymp} \& \ref{eq:Uwall_muS} for $\tilde{u}_w^{\infty}$ can now be verified by plotting the steady-state wall speed with respect to the normalised wall friction coefficient $\tilde{\mu}_w$ as in figure \ref{fig:Fig4_WallSpeedSS}. Here, the exact wall speed (open circles) is computed by substituting the steady state solution \ref{eq:MuSteadyState} in equation \ref{eq:IreWrittenNonDim} and integrating it numerically. The asymptotic results \ref{eq:Uwall_asymp} \& \ref{eq:Uwall_muS} are displayed as continuous (blue) and  dashed (red) lines, respectively. The normalised wall velocity varies slowly with the normalised friction coefficient $\tilde{\mu}_w$ as long as $1 - \tilde{\mu}_w$ is small (or $\mu_w \sim \mu_s$). Thus, for a given normal stress  $p_w$ at the wall, $\tilde{u}_w$ varies linearly with the applied shear stress $\tau_w = \mu_w p_w$ if the latter is sufficiently close to the yield shear stress $\tau_s = \mu_s p_w$. Then the wall velocity increases rapidly with $1 - \tilde{\mu}_w$ and it becomes logarithmically large as $\mu_w$ approaches $\mu_d$. This is not surprising since when $\tau_w$ approaches $\mu_d p_w$, the \textit{Inertial} number $I \gg 1$ in such a manner that frictional grain-grain contacts become less dominant compared to a grain-grain collisions as internal grain rearrangements are much frequent compared to the local deformation rate \citep{Andreotti_Forterre_Pouliquen_2011}. Thus, a highly agitated flow can occur near the wall. Indeed, beyond this critical value, there is no longer an equilibrium between the applied shear stress and the internal resistance via frictional contacts as previously pointed out \citet{cawthorn2011thesis}. A more relevant description is given by models inspired from kinetic theory of gases \citep{jenkins1983theory, goldhirsch2003rapid}. Furthermore, figure \ref{fig:Fig4_WallSpeedSS} indicates that the expression \ref{eq:Uwall_muS} provides a very good approximation to the steady-state wall speed when the applied shear stress is just above the yield stress $\tau_s = \mu_s p_w$. And the expression \ref{eq:Uwall_asymp} remarkably captures the wall speed variation for all applied shear stress $\tau_w = \mu_w p_w \in [\mu_s p_w, \mu_d p_w] $.

The above result that the wall speed $\tilde{u}_w(\tilde{t}) \propto \sqrt{\tilde{t}}$ along with the fact that $\tilde{u}_w(\tilde{t} \gg 1)$ follows an universal trend \ref{eq:Uwall_asymp} suggests that the velocity profile can be approximately deduced from the short-time asymptotic solution \ref{eq:SolutionSS}. This hypothesis is explored in figure \ref{fig:Fig5_VelocityProfile} which shows three different normalisations of the velocity field as a function of time. Each column corresponds to specific wall boundary conditions corresponding to $\tilde{\mu}_w = 0.9$ (left),  $\tilde{\mu}_w = 0.5$ (middle) and  $\tilde{\mu}_w = 0.1$ (right). The steady-state solution is displayed as a thick dashed line only in the first row but it is left out in the rest of the graphs for the sake of clarity. In this row, all figures present the time development of the velocity $\tilde{u}(\tilde{y}, \tilde{t})$ as computed by numerically integrating \ref{eq:IreWrittenNonDim} at each $\tilde{t} = 10^{-3}, 10^{-2}, 10^{-1}, 10^{0}, 10^{1}$. As seen before in figure \ref{fig:Fig3_WallSpeedComparison}, it can be readily observed that the wall velocity $\tilde{u}(\tilde{y}=0, \tilde{t})$ increases in time and attains a steady-state value which in turn depends on the applied wall shear stress. The numerical solution at $\tilde{t} = 10^{1}$ is already superposed on the steady state solution (dashed line) for all the cases of $\tilde{\mu}_w$ shown here. The second and the third row in figure \ref{fig:Fig5_VelocityProfile} display the same data when the velocity field is normalised with the wall speed $\tilde{u}_w(\tilde{t}) = \tilde{u}(\tilde{y}=0, \tilde{t})$ and $\tilde{u}_w^{\infty} \sqrt{\tilde{t}}$, respectively, as a function of $\tilde{y}/2 \Delta \mu \sqrt{\tilde{t}}$. Note that $\tilde{u}_w^{\infty}$ is taken from the expression \ref{eq:Uwall_asymp}. Irrespective of the velocity normalisation, all the velocity profiles collapse on to an unique curve expect for the cases when $\tilde{t} > \mathcal{O}(1)$ as expected from the previous results. However, the collapse is only marginally good when $\tilde{\mu}_w = 0.1$ when $\tilde{t} \sim \mathcal{O}(1)$ or greater. This implies that this observation may apply at shorter and shorter times as $\mu_w$ tends towards $\mu_d$. Nonetheless, as long as $\tilde{t} \leq \mathcal{O}(1)$, the velocity field should be given by $\tilde{u}(\tilde{y}, \tilde{t}) \sim \tilde{u}_w^{\infty}\sqrt{\tilde{t}} f\left( \frac{\tilde{y}}{2 \Delta \mu \sqrt{\tilde{t}}} \right)$, where $\tilde{u}_w^{\infty}$ is given by the expression \ref{eq:Uwall_asymp}.
%%%%%%%%%%%%%%%%%%%%%%%%%%%%%%%%%%%%%
%\begin{equation}
%\tilde{u}(\tilde{y}, \tilde{t}) \sim -\frac{\sqrt{\tilde{t}}}{\Delta \mu} \log \left[ \frac{1}{4} \left( \frac{\sqrt{\frac{\mu_w}{\mu_s} + 1}}{\sqrt{\frac{\mu_w}{\mu_s} - 1}} \right)  \beta_w \tilde{\mu}_w \right] f\left( \frac{\tilde{y}}{2 \Delta \mu \sqrt{\tilde{t}}} \right),
%\label{eq:Uprofile_muD}
%\end{equation}
%%%%%%%%%%%%%%%%%%%%%%%%%%%%%%%%%%%%%

%%%%%%%%%%%%%%%%%%%%%%%%%%%%%%%%%%%%
\subsection{Stokes's first problem with imposed wall velocity}
\label{subsec:ImposedVelocity}
So far, in this article Stokes' first problem is considered for the case when the wall shear stress $\tau_w$ is imposed externally. Therefore, the wall velocity developed with time and, as the internal resistance of the granular media balances out the applied stresses, it saturated to a constant value $u_w$. In contrast, it should be possible to set the granular media under motion by imposing the wall speed $u_w$. Here, the resulting shear stress experienced by the wall should vary temporally as the internal resistance of the granular media develops with time. However, it should also ultimately saturate to a constant value $\tau_w$ so that it gives precisely the same steady state solution as for the imposed shear stress case. In this subsection, a brief note on this variant of the Stokes' first problem is presented.

As already seen in figure \ref{fig:Fig4_WallSpeedSS}, in the steady state solution, for each wall friction coefficient there exists one and only one wall velocity. Therefore, it is reasonable to leave the normalised variables of the previous sections as such. Now by using the normalised local friction coefficient $\tilde{\mu} = (\mu_d - \mu)/\Delta \mu$, the equation \ref{eq:GranularStokesEqn_mu} can be rewritten as
%%%%%%%%%%%%%%%%%%%%%%%%%%%%%%%%%%%%
\begin{equation}
\frac{\partial \tilde{\mu}}{\partial \tilde{t}} = \Delta \mu^2 \tilde{\mu}^2 \left( \frac{2 \beta_w}{\sqrt{1 + \beta_w \tilde{y}}} \frac{\partial \tilde{\mu}}{\partial \tilde{y}} + \sqrt{1 + \beta_w \tilde{y}} \frac{\partial^2 \tilde{\mu}}{\partial \tilde{y}^2} \right),
\label{eq:GranularStokesEqn_muBIS}
\end{equation}
%%%%%%%%%%%%%%%%%%%%%%%%%%%%%%%%%%%%
which for the case of Stokes's first problem with imposed wall velocity should satisfy the initial condition $\tilde{\mu}(\tilde{y}, \tilde{t} = 0) = 1$ (or $\mu = \mu_s$) along with the boundary conditions
%%%%%%%%%%%%%%%%%%%%%%%%%%%%%%%%%%%%%%
\begin{eqnarray}
	\tilde{u}_w =  \int_{0}^{1} {\frac{\sqrt{\tilde{p}}}{\Delta \mu} \left( \frac{1}{\tilde{\mu}} - 1 \right)} d \tilde{y},
	%&& -\frac{2}{3 \beta_w} \left(\frac{1 + \tilde{\mu}_w}{\Delta \mu^2}\right) \left((1 + \beta_w)^{3/2} - 1 \right),
\label{eq:GranularStokesEqn_muBIS_BCs}
\end{eqnarray}
%%%%%%%%%%%%%%%%%%%%%%%%%%%%%%%%%%%%%
and $\tilde{\mu}(\tilde{y}=1, \tilde{t}) = 1$. Here, the initial and lower boundary conditions are chosen so as to satisfy $\tilde{u} = 0$ which is possible only when $\mu = \mu_s$ in the $\mu(I)$-rheology. The integral condition imposes the wall velocity on the choice of the vertical distribution of the normalised local friction coefficient $\tilde{\mu}$. In particular, note that the parameter $\beta_w = \phi \rho g \delta_s / p_w$ as in \ref
{eq:ShearLayerThickness} since the steady-state solution \ref{eq:MuSteadyState} for the Stokes' first problem with applied shear stress should also apply to this case where the wall speed is externally imposed. Thus, $\beta_w = \mu_w^{\infty}/\mu_s -1$ where $\mu_w^{\infty}$ is the steady state wall friction coefficient that is needed to sustain the applied wall speed $u_w$.

\begin{figure}
\begin{center}
\epsfig{file=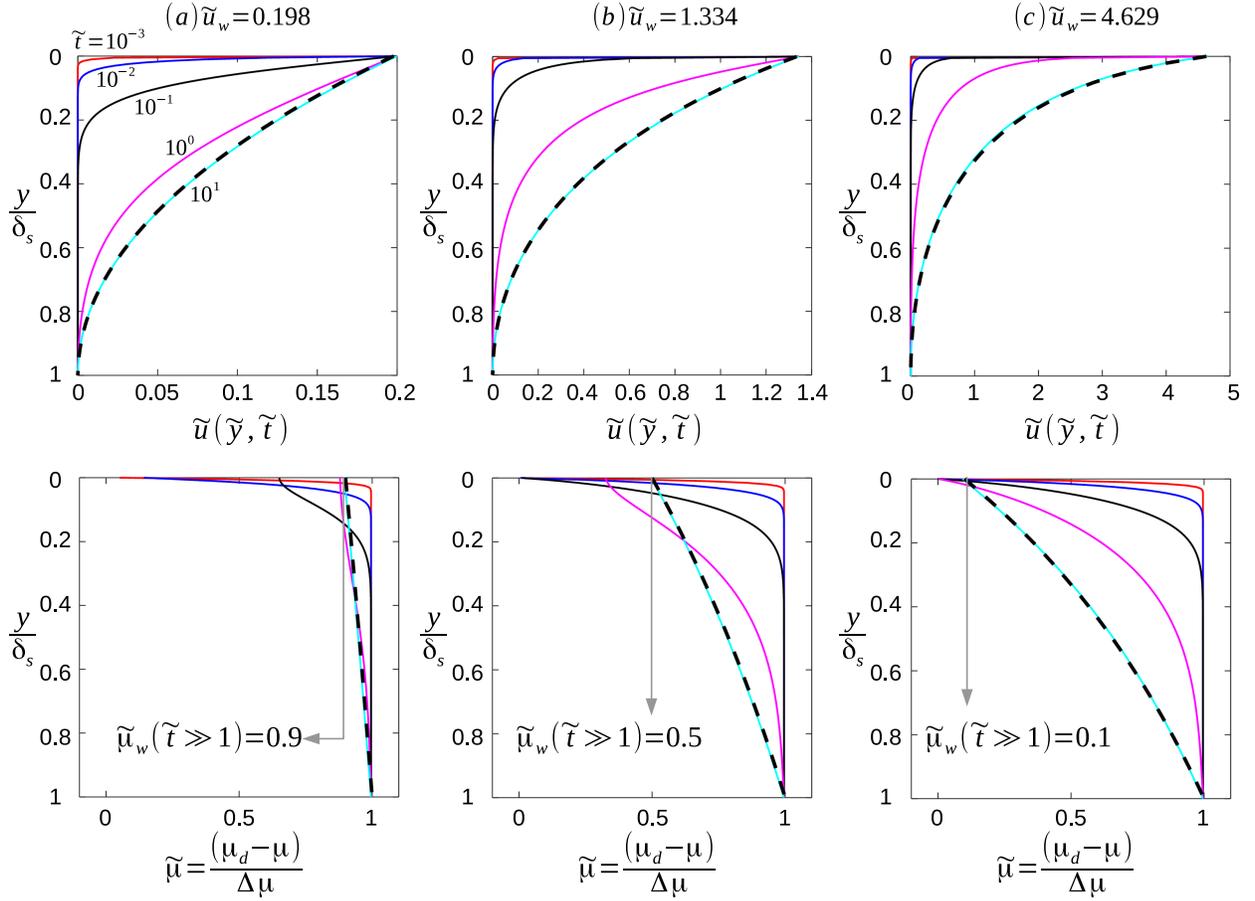,width=1\textwidth,keepaspectratio=true}
\end{center}
\caption{Numerical results for Stokes's first problem with imposed wall velocity showing the evolution of the velocity $\tilde{u}(\tilde{y}, \tilde{t})$ (top) and the normalised friction coefficient $\tilde{\mu}(\tilde{y}, \tilde{t})$ (bottom) at different time $\tilde{t} = 10^{-3}, 10^{-2}, 10^{-1}, 10^{0}, 10^{1}$ for various imposed wall speed $\tilde{u}_w$.  These data are obtained by directly integrating the numerical solution of the non-dimensional shear stress equation \ref{eq:GranularStokesEqn_muBIS} with the wall boundary condition \ref{eq:GranularStokesEqn_muBIS_BCs}. Thick dashed line represents the steady-state solution. The imposed wall speed $\tilde{u}_w$ had been chosen to match with the steady-state wall speed in figure \ref{fig:Fig5_VelocityProfile}.}
\label{fig:Fig6_VelocityAndMUProfile_WallSpeedImposed}
\end{figure}
\begin{figure}
\begin{center}
\epsfig{file=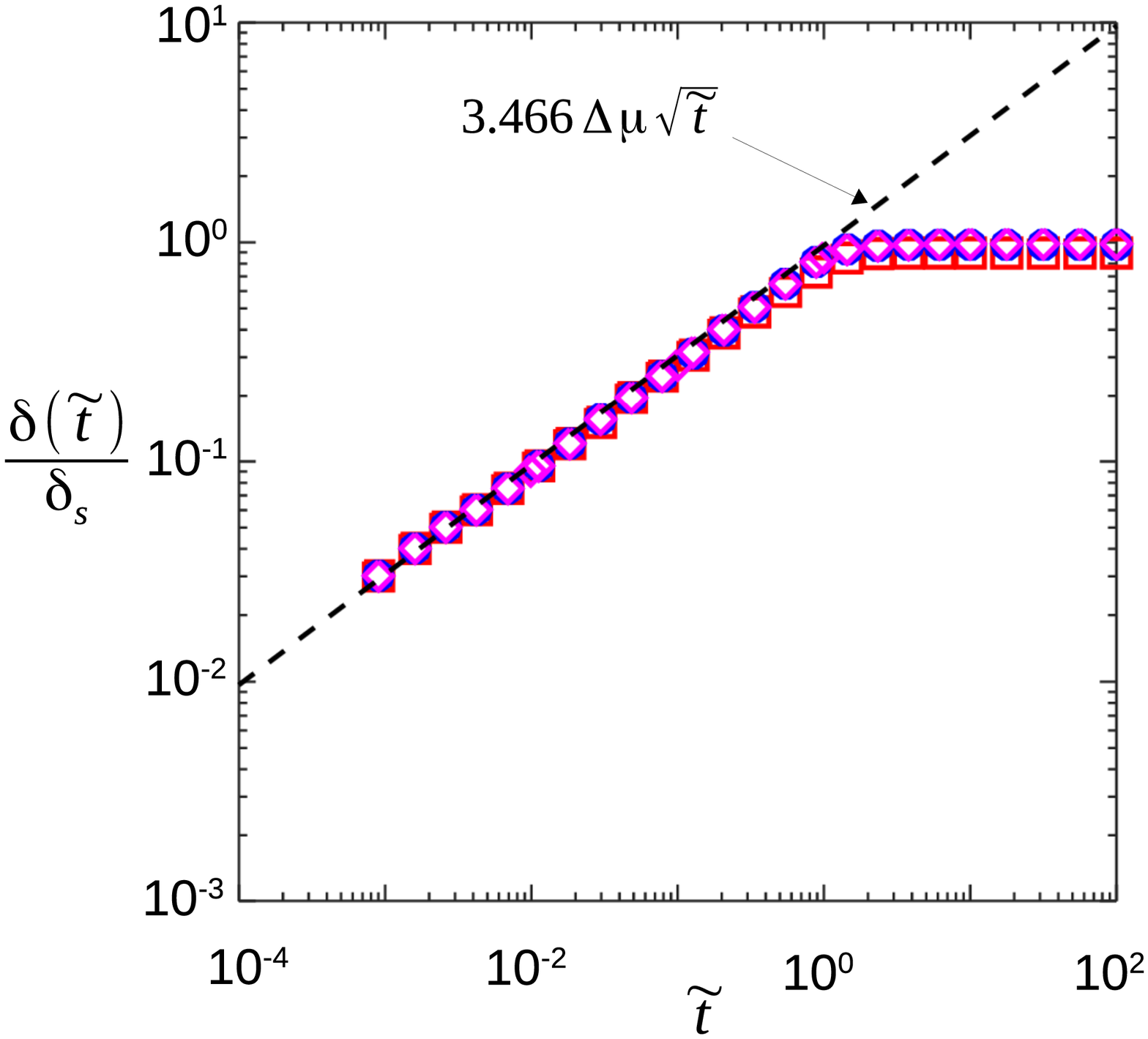,width=0.5\textwidth,keepaspectratio=true}
\end{center}
\caption{Temporal evolution of the shear layer thickness $\delta/ \delta_s$ for the profiles presented in figure \ref{fig:Fig6_VelocityAndMUProfile_WallSpeedImposed}. Different symbols correspond to various imposed wall velocity ($\square$ : $\tilde{u}_w = 0.198$, $\circ$ : $\tilde{u}_w = 1.334$ \& $\diamond$ : $\tilde{u}_w = 4.629$). As already seen in figure \ref{fig:Fig1_ShearStressNum} (bottom) for the case when the wall shear stress is imposed, the shear layer grows as $\sqrt{\tilde{t}}$ until about $\tilde{t} \sim \mathcal{O}(1)$.}
\label{fig:Fig7_SelfSimilarMU_VelocityImposed}
\end{figure}

Figure \ref{fig:Fig6_VelocityAndMUProfile_WallSpeedImposed} presents the velocity $\tilde{u}(\tilde{y}, \tilde{t})$ (top) and the normalised friction coefficient $\tilde{\mu}(\tilde{y}, \tilde{t})$ (bottom) profiles that are obtained by numerically solving eqn. \ref{eq:GranularStokesEqn_muBIS} for $\tilde{\mu}$ satisfying the imposed wall velocity condition as given by \ref{eq:GranularStokesEqn_muBIS_BCs}. Each continuous line represents different time as indicated in the figure $(\tilde{t} = 10^{-3}, 10^{-2}, 10^{-1}, 10^{0}, 10^{1})$ and thick dashed line represents the steady-state solution. Note that particular wall speeds ((a)$\tilde{u}_w = 0.198$, (b) $\tilde{u}_w = 1.334$ \& (c) $\tilde{u}_w = 4.629$) were chosen so that the resulting steady-state velocity and local friction coefficient profiles are the same as that obtained in figure \ref{fig:Fig5_VelocityProfile} for the case when wall shear stress is imposed. As expected, in all figures, the velocity profile (top) and the related boundary layer develops with time in such a way that the velocity at the wall is equal to the applied wall speed $\tilde{u}_w$ at all times and the size of the mobile layer increases until about $\tilde{t} \sim \mathcal{O}(1)$. This is true for the normalised wall friction coefficient (bottom) as well. Also, both $\tilde{u}(\tilde{y}, \tilde{t})$ and $\tilde{\mu}(\tilde{y}, \tilde{t})$ match with their respective steady-state profiles (thick dashed line) for sufficiently large time $\tilde{t} > \mathcal{O}(1)$.

\begin{figure}
\begin{center}
\epsfig{file=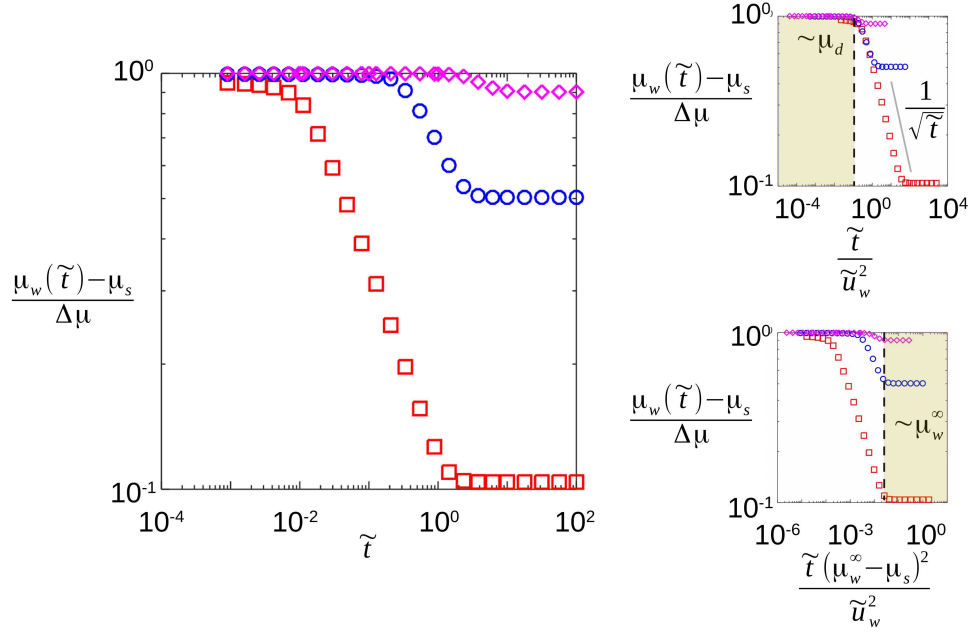,width=0.8\textwidth,keepaspectratio=true}
\end{center}
\caption{Evolution of the wall friction coefficient $\mu(\tilde{y} = 0, \tilde{t}) = \mu_w(\tilde{t})$ for various imposed wall velocity ($\square$ : $\tilde{u}_w = 0.198$, $\circ$ : $\tilde{u}_w = 1.334$ \& $\diamond$ : $\tilde{u}_w = 4.629$). Figures on the right display the same data as a function of two different normalisations of the time variable based on the scaling laws \ref{eq:TimeScaleMuD} \& \ref{eq:TimeScaleMuW}.}
\label{fig:Fig8_MUwall_VelocityImposed}
\end{figure}

With these data, it is then possible to calculate the size of the mobile layer, say $\delta(\tilde{t})$, where the local shear stress is just above the threshold shear $\tau_s = \mu_s p$. Figure \ref{fig:Fig7_SelfSimilarMU_VelocityImposed} illustrates that $\delta(\tilde{t})$ increases with time as $\sqrt{\tilde{t}}$ until $\tilde{t} \sim \mathcal{O}(1)$, after which it saturates to the steady-state limit. Even though this is similar to what was observed previously for the case of the Stokes' problem when the wall shear is imposed as seen in figure \ref{fig:Fig1_ShearStressNum} (bottom), the evolution of the wall friction coefficient $\mu_w(\tilde{t})$ takes place in two stage (see figure \ref{fig:Fig8_MUwall_VelocityImposed}). It is observed that $\mu_w(\tilde{t})$ is close to $\mu_d$ for sufficiently small time, say up to when $\tilde{t} \leq \tilde{t}_{\mu d}$. Then it decreases as $1/\sqrt{\tilde{t}}$ until it attains the steady-state at $\tilde{t} \sim \tilde{t}_{\mu w}$.  Thus, the corresponding wall shear stress $\tau_w(\tilde{t}) = \mu_w(\tilde{t}) p_w$ initially remains sufficiently close the critical shear stress $\mu_d p_w$ before decreasing monotonically towards the wall shear stress that is needed to sustain the applied wall velocity $\tilde{u}_w$. It suggests that the time at which these two stages occur should depend on $\tilde{u}_w$ and these timescales are different from $\delta_s^2/\nu_g$. 

For $\tilde{t}_{\mu d} \leq \tilde{t} \leq \tilde{t}_{\mu w}$, an order of magnitude analysis of eqn. \ref{eq:StokesEqn1} gives ${u_w}/{t} \sim {(\mu_w(t) - \mu_s)p_w}/{\delta(t)}$. In terms of the non-dimensional time $\tilde{t}$ and wall velocity $\tilde{u}_{w}$, this can be rewritten as
%%%%%%%%%%%%%%%%%%%%%%%%%%%%%%%%%%%%%%%
\begin{eqnarray}
	\mu_w(\tilde{t}) - \mu_s \sim \frac{\tilde{u}_w}{3.66 \Delta \mu \sqrt{t}},
\label{eq:TimeScaleMu}
\end{eqnarray}
%%%%%%%%%%%%%%%%%%%%%%%%%%%%%%%%%%%%%%
since $\delta \sim 3.66 \Delta \mu \sqrt{\tilde{t}}$ as shown by figure \ref{fig:Fig7_SelfSimilarMU_VelocityImposed}. Since when $\tilde{t} \sim \tilde{t}_{\mu d}$, $\mu_w(\tilde{t}) - \mu_s \approx \Delta \mu$ and this implies that the wall friction coefficient should be around $\mu_d$ until some time
%%%%%%%%%%%%%%%%%%%%%%%%%%%%%%%%%%%%%%%
\begin{eqnarray}
\tilde{t}_{\mu d} \propto \tilde{u}_w^2.
\label{eq:TimeScaleMuD}
\end{eqnarray}
%%%%%%%%%%%%%%%%%%%%%%%%%%%%%%%%%%%%%%
Thus, according to $\mu(I)$--rheology in order to sustain an applied wall speed $\tilde{u}_w$ with no slip condition, the granular media develops a strong shear stress (about $\mu_d p_w$) near the wall for all time $\tilde{t} \leq \tilde{t}_{\mu d}$. From there onwards, the wall shear stress should decrease monotonically as $1/\sqrt{\tilde{t}}$ until when the local friction coefficient (or the local shear stress) distribution has reached the stead-state solution to give $\mu_w(\tilde{t} \gg 1) = \mu_w^{\infty}$. Here, $\mu_w^{\infty}$ is a function of the applied wall speed and it can be either estimated directly by integrating the steady-state profile \ref{eq:MuSteadyState} or approximately via the expression \ref{eq:Uwall_asymp}. Using \ref{eq:TimeScaleMu}, it is seen that the wall friction coefficient and hence, the wall shear stress should attain a steady-state value at some time
%%%%%%%%%%%%%%%%%%%%%%%%%%%%%%%%%%%%%%%
\begin{eqnarray}
\tilde{t}_{\mu w} \propto \left(\frac{\tilde{u}_w}{\mu_w^{\infty} - \mu_s} \right)^2.
\label{eq:TimeScaleMuW}
\end{eqnarray}
%%%%%%%%%%%%%%%%%%%%%%%%%%%%%%%%%%%%%% 
These two timescales can be verified by plotting the data from figure \ref{fig:Fig8_MUwall_VelocityImposed} (left) as a function of properly rescaled time with respect to the relations \ref{eq:TimeScaleMuD} \& \ref{eq:TimeScaleMuW}. This is done in figure \ref{fig:Fig8_MUwall_VelocityImposed} (see plots on the right) where the data collapse indicates a good agreement with the above scaling laws.

\begin{figure}
\begin{center}
\epsfig{file=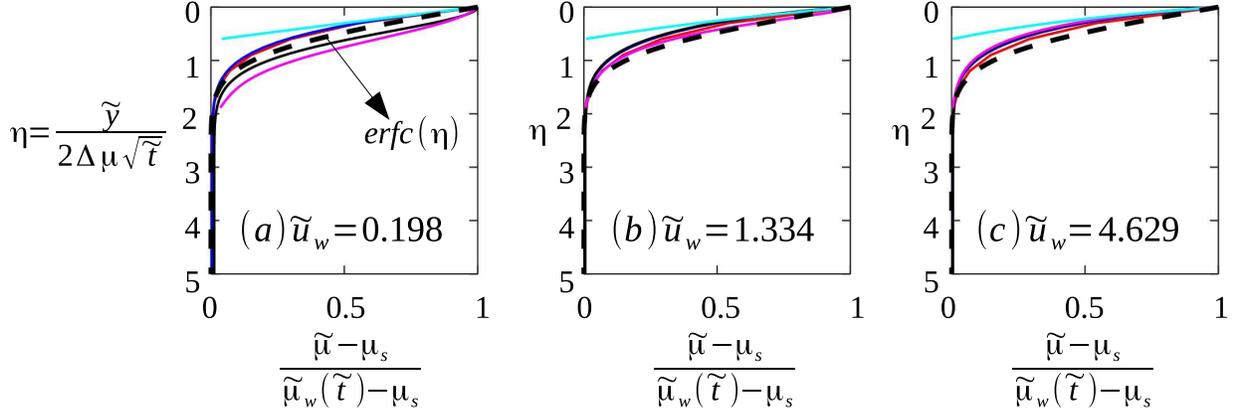,width=1\textwidth,keepaspectratio=true}
\end{center}
\caption{Same data as in figure \ref{fig:Fig6_VelocityAndMUProfile_WallSpeedImposed} (bottom) for different time $\tilde{t} = 10^{-3}, 10^{-2}, 10^{-1}, 10^{0}, 10^{1}$ but plotted here against the self-similar variable $\eta = \tilde{y}/2 \Delta \mu \sqrt{\tilde{t}}$. A reasonably good collapse is observed.}
\label{fig:FIG9_ShortTimeMuProfil_VelocityImposed}
\end{figure}
%%%%%%%%%%%%%%%%%%%%%%%%%%%%%%%%%%%
Finally, an approximate expression can now be elaborated for $\mu(\tilde{y},\tilde{t})$ due to the fact that the shear layer $\tilde{\delta}$ is small up to $\tilde{t} \sim \mathcal{O}(1)$. As done before in section \ref{subsec:SelfSimilarSolution}, the equation \ref{eq:GranularStokesEqn_muBIS} becomes
%%%%%%%%%%%%%%%%%%%%%%%%%%%%%%%%%%%%
\begin{equation}
\frac{\partial \tilde{\mu}}{\partial \tilde{t}} = \left( \Delta \mu^2 \tilde{\mu}^2 \right) \frac{\partial^2 \tilde{\mu}}{\partial \tilde{y}^2},
\label{eq:GranularStokesEqn_muBIS_ShortTime}
\end{equation}
%%%%%%%%%%%%%%%%%%%%%%%%%%%%%%%%%%%%
at short-time $\tilde{t}$. This equation is singular near the wall region $\tilde{y} \ll 1$ since $\tilde{\mu} \sim 0$ at least until $\tilde{t}_{\mu d}$ (see figure \ref{fig:Fig8_MUwall_VelocityImposed}). Nonetheless, the chosen initial condition $\tau (y, t = 0) = \mu_s p$ implies that there exists a zone where $\mu \sim \mu_s$ (or $\tilde{\mu} \sim 1$) away from the wall until the shear layer is completely developed i.e. for all $\tilde{t} \leq \mathcal{O}(1)$. In this \textit{outer} zone $\Delta \mu \tilde{\mu}\sim \Delta \mu$ and hence, the above equation reduces to a simple diffusion equation wherein the \textit{outer} solution should be $\mu = \mu_s + A \mbox{ erfc}(\tilde{y}/2 \Delta \sqrt{t})$. Here $A$ is an arbitrary constant that could be deduced by matching this solution to the \textit{inner} region where the friction coefficient $\tilde{\mu} \sim 0$. Here, $\mu \sim \mu_w (\tilde{t})$ and therefore
%%%%%%%%%%%%%%%%%%%%%%%%%%%%%%%%%%%%
\begin{equation}
\mu (\tilde{y}, \tilde{t}) = \mu_s + (\mu_w(\tilde{t}) - \mu_s) \mbox{ erfc}\left(\frac{\tilde{y}}{2 \Delta \mu \sqrt{\tilde{t}}}\right),
\label{eq:muShortTime_VelocityImposed}
\end{equation}
%%%%%%%%%%%%%%%%%%%%%%%%%%%%%%%%%%%%
is obtained as an approximate expression for the friction coefficient $\mu (\tilde{y}, \tilde{t})$ as long as $\tilde{t}$ is sufficiently small. When this expression is compared with the numerical results (see figure \ref{fig:FIG9_ShortTimeMuProfil_VelocityImposed}) a reasonable fit is observed for all $\tilde{t} \leq \mathcal{O}(1)$. As previously shown from computational data in figure \ref{fig:Fig8_MUwall_VelocityImposed}, it is pointed out here that $\mu_w(\tilde{t}) - \mu_s \sim \Delta \mu$ for all $\tilde{t} \leq \tilde{t}_{\mu d}$ and $\mu_w(\tilde{t}) - \mu_s \sim \tilde{u}_w/\Delta \mu \sqrt{\tilde{t}}$ when $\tilde{t}_{\mu d} \leq \tilde{t} \leq \tilde{t}_{\mu w}$.

\section{Conclusion}
\label{sec:Conclu}
Using the $\mu(I)$--rheology, the so-called Stokes' first problem on the motion of a granular liquid set-in by an impulsively started flat plate is studied both numerically and theoretically. The problem was first well-posed in terms of a non-linear diffusion equation for the local shear stress with proper initial and boundary conditions in order to avoid Hadamard instability. Numerical solutions are then obtained for both externally imposed wall stress and speed. Approximate solutions at short and long times are also illustrated to capture the main features of the numerical results.

For the case when the dense granular flow is brought about by applying constant shear stress $\tau_w$ at $t > 0$, if $\tau_w$ is greater than the yield stress at the wall $\mu_s p_w$ (where $p_w$ is the applied pressure at the wall) then the $\mu(I)$--rheology implies that it is diffused into the granular media until the shear stress is uniform throughout the media. Such that, at any time $t$, the applied shear stress reaches a depth proportional to $\sqrt{\nu_g t}$ where $\nu_g = (d/2\phi \Delta \mu I_0) \sqrt{p_w/\rho}$ denotes the diffusion coefficient for the local shear stress. A steady-state, wherein a finite zone of grains (of thickness, say, $\delta_s$) yield, and hence flow, due to the applied shear, is thus shown to occur at about a finite time of the order $\delta_s^2/\nu_g$. Here, if $\beta_w = (\mu_w - \mu_s)/\mu_s$, the shear layer depth $\delta_s$ is $\beta_w p_w / \phi \rho g$ as already obtained by \citet{cawthorn2011thesis, Andreotti_Forterre_Pouliquen_2011}. If no-slip condition is allowed, the wall velocity develops in time as $u_w^{\infty} \sqrt{\nu_g t}/ \delta_s$ until about a time $t \sim \mathcal{O}(\delta_s^2/\nu_g)$.

If the dense granular flow is set-up by suddenly imparting a constant speed $u_w$ on the wall at $t > 0$, the internal resistance of the media develops in a small region close to the wall and later it is diffused into the shear layer. As a result, the $\mu(I)$--rheology suggests that, initially when the wall is set into motion, the shear stress experienced by the granular media in the neighbourhood of the velocity driven wall should be sufficiently close to the critical shear stress $\tau_d = \mu_d p_w$ until some time $t_{\mu d}$ proportional to $\nu_g (u_w/g \delta_s)^2$. Thereafter, the wall shear stress $\tau_w$ decreases with time as a power law $1/\sqrt{t}$ before reaching its steady state value necessary to support the externally imposed wall speed $u_w$. At this stage the shear stress becomes uniform in the bulk of the mobile layer.

In both variants of Stokes' problem in granular media, a properly rescaled friction coefficient (or the shear stress) is illustrated to be approximately self-similar with respect to the variable $y/\Delta \mu \sqrt{\nu_g t}$. Moreover, the steady-state wall speed, wall shear stress and the applied wall pressure are related by a simple approximate expression \ref{eq:Uwall_asymp} which in terms of dimensional parameters can be given as $u_w \sim ({g \delta_s^2}/{\nu_g}) f( \beta_w ) $ where $f(\beta_w)$ is a function of the \textit{surplus} steady-state wall shear stress $\beta_w = (\tau_w - \mu_s p_w)/\mu_s p_w$ such that it is either $\mathcal{O}(1)$ when the wall shear stress is just above the yield stress $\tau_w \sim \tau_s$ or logarithmically large as $\tau_w$ approaches $\tau_d$.

Note that when the local friction coefficient $\mu$ approaches $\mu_d$, the local \textit{Inertial} number which compares the timescale of grain-grain rearrangements with that arising from the macroscopic deformation of the media should be very large ($I \gg 1$). Therefore, the aforementioned result that, for $t \leq t_{\mu d}$, the local shear stress $\tau = \mu p$ tends towards $\tau_d = \mu_d p$ in the immediate vicinity of the region where velocity is imposed, suggests that a highly agitated granular flow could occur in this zone. A viscoplastic description of the $\mu(I)$--rheology might, in fact, be not suitable in such zones  as the local shear stress therein cannot be supported by internal grain-grain frictional resistance alone. Here, a proper model for rapid granular flows \citep{jenkins1983theory, goldhirsch2003rapid} should be more pertinent. Furthermore, if the applied wall speed $u_w$ is much larger than $\nu_g / g \delta_s^2$ so that $t_{\mu d}$ becomes sufficiently large, the resulting unsteady granular flow as computed from the $\mu(I)$--rheology may even be incorrect.

It is expected that this study motivates investigations on the further validity of the $\mu(I)$--rheology for unsteady dense granular flows using simple experiments. These results should also be helpful to better understand shear layers, effective viscosity, drag force and characteristic diffusion timescales in future studies with the $\mu(I)$--rheology. Especially in the context of ill-posedness of the $\mu(I)$--rheology as an initial-value problem, it might be essential to identify what features predicted by this rheology are still meaningful. It will be of some interest to include the spatio-temporal variation of the solid fraction $\phi$ as well -- via, for example, a linear function of the inertial number $I$ as in \citet{jop2006constitutive}. Non-local effects by which a granular media can yield even if the local shear stress is below the yield criterion are omitted in the present short note for sake of simplicity but they might play an important role under common experimental conditions. Finally, it is pointed out that the Stokes' $2$nd problem with an oscillating wall boundary condition remains a very interesting open problem as it might shed light upon how static and dynamic zones can simultaneously appear and move around in unsteady flow fields predicted by $\mu(I)$--rheology of dense granular flows. However, the $\mu(I)$--rheology can neither account for the history of the shear stress in the bulk of the media nor consider other static initial conditions that are different from the yield criterion. It is nonetheless important to study these configurations using the $\mu(I)$--rheology to further advance knowledge about continuum models for unsteady dense granular flows.

% remerciements	
The authors acknowledge vital inputs from Simon Dagois Bohy and all correspondences with Delphine Doppler and Pierre Jop. %JJSJ also acknowledges all discussions with  Yo\"{e}l Forterre, Pierre-Yves Lagr\'{e}e and Olivier Pouliquen.
%%%%%%%%%%%%%%%%%%%%%%%%%%%%%%%%%%%

%\bibliographystyle{jfm}
%\bibliography{NLShearStressDiffusionInDenseGranularMedia}

\end{document}